\documentclass[prb,aps,graphicx,color,float]{revtex4}
\usepackage{epsfig}
\usepackage{bm}

\begin{document}

\preprint{}

\newcommand{\stef}[2]{$\blacktriangleright${\sc round #1:} {\em #2}$\blacktriangleleft$}

\title{Electronic Mach-Zehnder interferometer as a tool to probe fractional statistics}

\author{K. T. Law and D. E. Feldman}
\affiliation{Department of Physics, Brown University, 
Providence, RI 02912}
\author{Yuval Gefen}
\affiliation{Department of Condensed Matter Physics, Weizmann Institute of Science, Rehovot,
76100 Israel}

\begin{abstract}

We study transport through an electronic Mach-Zehnder interferometer recently devised at the Weizmann Institute.
We show that this device can be used to probe statistics of quasiparticles in the fractional quantum Hall regime.
We calculate the tunneling current through the interferometer as the function of the Aharonov-Bohm flux, temperature and voltage bias, and demonstrate that its flux-dependent component is strongly sensitive to the statistics of tunneling quasiparticles. 
More specifically, the flux-dependent and flux-independent contributions to the current are related by a power law, the exponent being a function of the quasiparticle statistics.

\end{abstract}


\maketitle

\section{Introduction}

One of the key features of the quantum Hall effect (QHE) is the fractional charge and statistics of quasiparticles. 
The seminal shot noise experiments of the Weizmann and Saclay groups \cite{weizmann, Grenoble} allowed for direct observations of fractional charges. 
A recent experiment on the mutual fractional statistics of two quasiparticles with different charges
has been published in Ref. \onlinecite{goldman2005}. It involved a setup consisting of an island of a fractional quantum Hall liquid embedded in a liquid with a different filling factor \cite{JKT}.
Theoretically, several approaches \cite{cfksw,HBT,kane} have been proposed as possible tools to probe mutual statistics of identical quasiparticles. However, to this date, no experimental verification of the statistics of identical quasiparticles has been reported.

In this paper we propose a different approach to observing the statistics of identical quasiparticles.
Our approach employs the electronic analogue of the Mach-Zehnder interferometer,
recently designed at the Weizmann Institute 
\cite{MZ}. This device has been used to observe the Aharonov-Bohm effect in the integer quantum Hall regime 
\cite{MZ}. 
As is shown below, in higher magnetic fields this type of device would allow the observation of fractional statistics.

The proposed method of the observation of fractional statistics has a number of advantages 
in comparison with other set-ups.
In Refs. \onlinecite{HBT,kane} one needs to measure the noise or the current-current correlation function.
In our approach it is sufficient to find the current through the interferometer. This resembles Ref. \onlinecite{cfksw}. However, in Ref. \onlinecite{cfksw} one has to control the number of quasiparticles trapped in the interferometer in order to probe fractional statistics.
There is no such difficulty in our case. Besides, the interference pattern can be observed in the Mach-Zehnder interferometer at a larger interferometer size than in the standard geometry \cite{cfksw}.

An electronic Mach-Zehnder interferometer is sketched in Fig. 1. Charge propagates along two quantum Hall edges and tunnels between the edges at the point contacts QPC1 and QPC2. Figs. 1a) and 1b) depict two possible setups: 
a) tunneling takes place between two different fractional quantum Hall puddles; and b) tunneling is between the edges of a single puddle. 
In the latter case, fractionally charged quasiparticles can tunnel at QPC1 and QPC2. In the former case, only electrons are allowed to tunnel \cite{wen}. Pinching off the edges closer to each other, one can deform the system depicted in Fig. 1b) into the configuration 
Fig. 1a).
The tunneling current between the edges depends on their voltage difference and the magnetic flux through the region A-QPC1-B-QPC2-A. As shown below, the current includes a flux-independent contribution $I_0$ and a contribution $I_\Phi$ which oscillates as function of the magnetic flux with period $\Phi_0=hc/e$. 
We will calculate $I_\Phi$ and $I_0$ as functions of tunneling amplitudes $\Gamma_1$
and $\Gamma_2$ at QPC1 and QPC2. If no tunneling occurs at QPC2, i.e. $\Gamma_2=0$, then the total current $I(\Gamma_1,\Gamma_2)=I_0(\Gamma_1,0)$ is independent of the magnetic flux. At weak tunneling at QPC2, i.e. small $\Gamma_2\ll\Gamma_1$, the flux-dependent component of the current scales as
$I_\Phi(\Gamma_1,\Gamma_2)\sim[I_0(\Gamma_1,\Gamma_2)-I_0(\Gamma_1,0)]^b$, where the exponent $b$ depends on the statistics of tunneling particles. In the case of electron tunneling, $b=1/2$. In the case of the tunneling of fractional quasiparticles, the exponent depends on their statistics and is always greater than for electrons.

The outline of the paper is the following. First, we give a qualitative explanation of our results.
Next, we introduce the model which is used in our calculations. Then, we calculate the $I-V$ curves, respectively, for quasiparticle tunneling in 
the fractional QHE regime
with filling factors $\nu=1/(2m+1)$, and electron tunneling between quantum Hall liquids with filling factors $\nu_{1,2}=1/(2m_{1,2}+1)$. In the latter case our calculations follow the standard route \cite{cfksw,geller}. In the former case one has to carefully treat the Klein factors describing quasiparticle statistics. 
The Appendices
 discuss some technical details of the perturbation theory employed.

\section{Qualitative discussion}

We study charge transport through the Mach-Zehnder interferometer in the limit of weak tunneling at QPC1 and QPC2. Thus, we assume that the edges of the fractional quantum Hall liquid are far from each other in comparison with the magnetic length $l_0$. Hence, $\Gamma_1,\Gamma_2\ll \hbar/\tau_c$, where the ultra-violet cut-off scale $\tau_c\sim l_0/v$ and $v$ is the excitation velocity along the edges. The tunneling amplitudes can be controlled by gate voltages. We assume that the gate voltages are chosen such that one of the situations shown in Figs. 1a) and 1b) takes place. Other situations, e.g. weak quasiparticle tunneling at QPC1 and weak electron tunneling at QPC2, are also possible but will not be considered here.

We first address the electron tunneling case depicted in Fig. 1a). There are three contributions to the current $I=I_1+I_2+I_{12}$. The first contribution is due to the tunneling at QPC1, $I_1=c_1|\Gamma_1|^2$, where $c_1$ depends on the temperature and voltage and is independent of the magnetic flux through the interferometer. The second contribution is due to the second quantum point contact, $I_2=c_1|\Gamma_2|^2$. Finally, a contribution arises due to quantum interference between the electrons which tunnel at QPC1 and QPC2. This contribution equals $I_{12}=c_2\Gamma_1\Gamma_2^*\exp(-i\phi)+c.c.$, where $c_2$ depends on the voltage, temperature and the distance between the point contacts,
and $\phi$ is the difference of the Aharonov-Bohm phases picked up by electrons propagating between QPC1 and QPC2 along two edges. The phase 
$\phi=\frac{e}{\hbar c}[\int_{\rm QPC1-A-QPC2}\vec A d\vec l-\int_{\rm QPC1-B-QPC2}\vec A d\vec l]=2\pi\Phi/\Phi_0$, where $\vec A$ is the vector potential,
$\Phi<0$ the magnetic flux through the region A-QPC2-B-QPC1-A, $e<0$ the electron charge,
and $\Phi_0$ the magnetic flux quantum $hc/e$. The coefficients $c_1$ and $c_2$ will be calculated in section IV. Thus, the current exhibits a periodic dependence on the magnetic flux $\Phi$ with period $\Phi_0$. The amplitude of the flux-dependent contribution to the current $I_\Phi={\rm max~}I_{12}$ is related to the flux-independent contribution $I_0=I_1+I_2$ by the equation

\begin{equation}
\label{s2e1}
I_{\Phi}(\Gamma_1,\Gamma_2)\sim[I_0(\Gamma_1,\Gamma_2)-I_0(\Gamma_1,0)]^{1/2}.
\end{equation}
Eq. (\ref{s2e1}) is derived rigorously in section IV.

A similar relation with a different exponent can be obtained for the set-up Fig. 1b) in which fractionally charged quasiparticles tunnel between the edges. Quasiparticles in a quantum Hall liquid with the filling factor $\nu=1/(2m+1)$ can be described as point charges $q=\nu e$ with attached solenoids \cite{anyons}. 
Each solenoid carries one magnetic flux quantum $\Phi_0$. When one quasiparticle makes a circle around another it picks up the Aharonov-Bohm phase

\begin{equation}
\label{s2e2}
\theta=2\pi \nu
\end{equation}
which describes fractional statistics. The total flux $\tilde\Phi$ through the interferometer includes the contribution from the applied magnetic field $\Phi=\int \vec B d\vec S$ and the statistical contribution from the 
flux tubes attached to quasiparticles.

Fig. 2 is topologically equivalent to Fig. 1b).  As is clear from Fig. 2, the solenoids attached to quasiparticles do not contribute to the magnetic flux through the dashed circle, if there is no tunneling between edges 1q and 2q. Thus, in the absence of tunneling the total magnetic flux $\tilde\Phi=\Phi$. Each tunneling event changes the flux $\tilde\Phi$ by one flux quantum. The flux decreases by $|\Phi_0|$ when a quasiparticle tunnels from the outer edge to the internal edge, and 
increases by $|\Phi_0|$ for the tunneling events from edge 2q to edge 1q. Hence, $\tilde\Phi=\Phi+n\Phi_0$, where $n$ is an integer.


We will assume that the electrochemical potential of source S1 is higher than the electrochemical potential of source S2 by $eV$, where $e$ is an electron charge. These electrochemical potentials are equal to the chemical potentials of edges 1q and 2q. 


Let us first consider the simplest limit of zero temperature. In this case, quasiparticles can tunnel from the edge with the higher chemical potential to the edge with the lower chemical potential only and cannot tunnel from edge 2q to edge 1q. The tunneling probability can be derived in the same way as Eq. (\ref{s2e1}),

\begin{equation}
\label{s2e3}
p(\tilde\Phi)=\tilde c_1 (|\Gamma_1|^2+|\Gamma_2|^2)+(\tilde c_2 \Gamma_1\Gamma^*_2\exp(-i\tilde\phi)+c.c.),
\end{equation}    
where $\tilde\phi=2\pi\nu\tilde\Phi/\Phi_0=2\pi\nu\Phi/\Phi_0+2\pi\nu n$ is the Aharonov-Bohm phase accumulated by a quasiparticle with the charge $\nu e$ along the path A-QPC2-B-QPC1-A. The phase $\tilde\phi$ is periodic in $n$ with period $1/\nu$. Let us assume that initially $n=k\time 1/\nu$. Then $p(\tilde\Phi)=p(\Phi)$ and the transfer of a quasiparticle requires the time $t_1=1/p(\Phi)$. After a quasiparticle tunneling event the statistical flux changes and the second quasiparticle tunnels after the  time interval $t_2=1/p(\Phi+\Phi_0)$. After $1/\nu$ tunneling events we return to the situation with $n$ being a multiple of $1/\nu$.
The time needed for the transfer of $1/\nu$ quasiparticles, i.e. a single electron charge, is $t=\sum_{l=1}^{1/\nu} t_l$. Hence, the current

\begin{equation}
\label{s2e4}
I=\frac{e}{t}=\frac{1/\nu}{\sum_{l=1}^{1/\nu}\frac{1}{I_l}},
\end{equation}
where $I_l=\nu e p(\Phi+[l-1]\Phi_0)$. If the quasiparticles did not carry flux tubes the current through the Mach-Zehnder interferometer in the presence of the magnetic flux $(\Phi+[l-1]\Phi_0)$ would equal $I_l$. Thus, the current of quasiparticles obeying fractional statistics is a harmonic average of $1/\nu$ currents corresponding to the imaginary situation of fractionally charged quasiparticles which do not obey fractional statistics.

If $\Gamma_2\ll\Gamma_1$ then the current $I$ can be expanded in powers of $\Gamma_2$.
The term of order $\Gamma_2^u\Gamma_2^{*w}$ is proportional to $\exp(2\pi\nu i[u-w]\Phi/\Phi_0)$. The current (\ref{s2e4}) is a periodic function of the magnetic flux with period $\Phi_0$. Hence, $[u-w]=k/\nu$ in all non-zero contributions.
Thus, the linear in $|\Gamma_2|$ contribution 
to the current vanishes. Hence, the leading power of $\Gamma_2$ in the expansion is $|\Gamma_2|^2$. Such second order terms do not exhibit a magnetic flux dependence. Hence, the flux-independent contribution to the current $I_0$ satisfies the equation

\begin{equation}
\label{s2e5}
I_0(\Gamma_1,\Gamma_2)-I_0(\Gamma_1,0)\sim |\Gamma_2|^2.
\end{equation}
The leading flux-dependent contribution scales as

\begin{equation}
\label{s2e6}
I_\Phi\sim {\rm const} \Gamma_2^{1/\nu}\exp(2\pi i\Phi/\Phi_0)+ c.c.
\end{equation}
Finally, one gets from the comparison of Eqs. (\ref{s2e5}) and (\ref{s2e6}) at $\Gamma_2\ll\Gamma_1$

\begin{equation}
\label{s2e7}
I_\Phi(\Gamma_1,\Gamma_2)\sim[I_0(\Gamma_1,\Gamma_2)-I_0(\Gamma_1,0)]^{\pi/\theta},
\end{equation} 
where $\theta$ is the statistical angle (\ref{s2e2}). Eq. (\ref{s2e7}) is derived rigorously in section V.
The flux-dependent and flux-independent contributions to the current can be expressed via the maximal and 
minimal values of the current as function of the magnetic flux: 
$I_0=[{\rm max}I+{\rm min}I]/2$, $I_{\Phi}=[{\rm max}I-{\rm min}I]/2$.
The exponent $\pi/\theta$ is determined by the quasiparticle statistics. It equals $1/2$ for fermions, Eq. (\ref{s2e1}),
and exceeds $1/2$ for fractional quasiparticles. 

Luttinger liquid effects are know to give rise to power laws in the edge physics in quantum Hall systems \cite{KF}. We would like to emphasize that Luttinger liquid physics is irrelevant for the above result, Eq. (\ref{s2e7}). The exponent $\pi/\theta$ emerges due to the statistical magnetic flux carried by quasiparticles, i.e. due to their fractional statistics.

We now discuss the finite temperature regime. In this case quasiparticles can tunnel both from edge 1q to edge 2q and
from edge 2q to edge 1q. The tunneling probabilities are related by the principle of detailed balance

\begin{equation}
\label{s2e8}
p_-(\Phi+[k\times\frac{1}{\nu}+r]\Phi_0)=\gamma p_+(\Phi+[k\times\frac{1}{\nu}+r-1]\Phi_0),
\end{equation}
where  
$\gamma=\exp(-\nu e V/k_B T)$;
$1\le r\le\nu$; the convention $r-1=1/\nu$ is used at $r=1$;
 $p_{+}(\tilde\Phi=\Phi~+{\bf~statistical ~flux ~before ~tunneling})$ and 
$p_-(\tilde\Phi=\Phi~+{\bf ~statistical ~flux ~before ~tunneling})$ 
denote the probabilities of the tunneling events from edge 1q to 2q and from edge 2q to edge 1q respectively. 
They depend quadratically on $\Gamma_{1,2}$ and are calculated in section V.
The tunneling probabilities depend on the total magnetic flux which includes the statistical contribution 
$\Phi_s=[k\times\frac{1}{\nu}+s]\Phi_0$, $1\le s\le 1/\nu$. 
Since the Aharonov-Bohm phase due to the statistical flux $\Phi_s$ equals
$\phi_{AB}=(2\pi k + 2\pi\nu s)$, the probabilities depend on $\Phi$ and $s$ only and are independent of $k$. Thus, the tunneling current is given by the equation

\begin{equation}
\label{s2e9}
I=e\nu \sum_{r=1}^{1/\nu} f_r[p_+(\Phi+r\Phi_0)-p_-(\Phi+r\Phi_0)],
\end{equation}
where $f_r$ is the probability to find the system in one of the states with $\Phi_s=[k\times\frac{1}{\nu}+r]\Phi_0$,
$1\le r\le1/\nu$ ,
$k$ being an arbitrary integer. The distribution function $f_r$ can be determined from the steady state condition

\begin{equation}
\label{s2e10}
f_r[p_+(\Phi+r\Phi_0)+p_-(\Phi+r\Phi_0)]=f_{r-1}p_+(\Phi+[r-1]\Phi_0)+f_{r+1}p_-(\Phi+[r+1]\Phi_0)],
\end{equation}
where we use the convention $1/\nu+1=1$. Using Eq. (\ref{s2e8}) the kinetic equation (\ref{s2e10}) can be rewritten as

\begin{equation}
\label{s2e11}
p_+(\Phi+r\Phi_0)[f_r - \gamma f_{r+1}]=p_+(\Phi+[r-1]\Phi_0)[f_{r-1} - \gamma f_r]; r=1,\dots, 1/\nu.
\end{equation}
One finds from the above system of equations

\begin{equation}
\label{s2e12}
f_r - \gamma f_{r+1}=\frac{\alpha}{p_+(\Phi+r\Phi_0)}, 
\end{equation}
where $\alpha$ is a constant independent of $r$. To calculate this constant we add Eqs. (\ref{s2e12}) with all possible 
$r$. Since $\sum f_r=1$, one gets

\begin{equation}
\label{s2e13}
\alpha=\frac{1-\gamma}{\sum_{r=1}^{1/\nu}\frac{1}{p_+(\Phi+r\Phi_0)}}.
\end{equation}
Thus, the current, Eq. (\ref{s2e9}), equals

\begin{equation}
\label{s2e14}
I=e\nu\sum_{r=1}^{1/\nu} p_+(\Phi+r\Phi_0)[f_r - \gamma f_{r+1}]=e\alpha=\frac{1/\nu}{\sum_{r=1}^{1/\nu}\frac{1}{I_r'}},
\end{equation}
where $I_r'=\nu e (1-\gamma) p_+(\Phi+r\Phi_0)$. Similarly to $I_l$, Eq. (\ref{s2e4}), the currents $I_r'$ can be understood as the currents of fictitious fractionally charged quasiparticles which do not obey fractional statistics.
A rigorous derivation of Eq. (\ref{s2e14}) is discussed in section V. Finally, the same analysis as in the zero-temperature case shows that Eq. (\ref{s2e7}) is satisfied at finite temperatures.

Eq. (\ref{s2e7}) is the main result of the article. An experimental test of this relation will allow the observation of fractional statistics. In such experiment one needs to change $\Gamma_2$ at fixed $\Gamma_1$. The tunneling amplitudes
$\Gamma_{1,2}$ are controlled by gate voltages. Generally, any change of gate voltages affects both tunneling amplitudes.
They can be controlled independently only if QPC1 and QPC2 are far from each other. In most interferometer set-ups
an increase in the distance between QPC1 and QPC2 would result in the suppression of the interference pattern. Fortunately, this is not the case for the Mach-Zehnder interferometer. The calculations for the electronic Mach-Zehnder interferometer in the integer quantum Hall regime \cite{FMZ1,FMZ2} show that the visibility of the interference pattern
depends not on the distance between QPC1 and QPC2 but only on the difference of the distances between the point contacts along two edges. We confirm the same conclusion for the fractional quantum Hall systems in sections IV and V.

In order to determine the relation between  the flux-dependent and flux-independent components of the current one has to vary the magnetic field and measure the current at different values of the field. 
The magnetic flux $\Phi$ through the region QPC1-A-QPC2-B-QPC1, Fig. 1b), includes two contribution: the flux through the hole in the interferometer, i.e. the upper half of the region QPC1-A-QPC2-B-QPC1, and the flux through
the lower half of the region QPC1-A-QPC2-B-QPC1 which is occupied by a quantum Hall liquid. If the magnetic field changes at the fixed density then the filling factor in the lower half deviates from $\nu$. As a result, quasiparticles can enter the region  QPC1-A-QPC2-B-QPC1. Each of them brings one flux quantum. This does not change any of the results of the paper.
Indeed, we predict that the current is a periodic function of the magnetic flux through the region  QPC1-A-QPC2-B-QPC1 with period $\Phi_0$. Hence, changing the flux by one flux quantum does not affect the current.

Several other quantum Hall interferometer set-ups have been discussed in the literature. The simplest set-up \cite{cfksw}
is illustrated in Fig. 3. In contrast to the Mach-Zehnder interferometer, the effective magnetic flux perceived by quasiparticles does not change after tunneling events in the set-up Fig. 3. Hence, the current is independent of the statistical phase $\theta$,
Eq. (\ref{s2e2}) (if no quasiparticles are trapped between the quantum point contacts). On the other hand, 
the current exhibits a ``fractional" Aharonov-Bohm periodicity with period $\Phi_0/\nu$. 
On the technical level the set-up \cite{cfksw} and our problem are described by very similar models
(see section III). The main difference consists in the Klein factors which describe fractional statistics. They are present in the model of the Mach-Zehnder interferometer  (section III) and are absent in the model of Ref. 
\onlinecite{cfksw}.
This difference between the models results in qualitatively different transport behavior.
A set-up related to the Mach-Zehnder interferometer was studied in Ref. \onlinecite{kane}, Fig. 4. The interferometer, Fig. 4, has the same topology as the Mach-Zehnder interferometer but includes three edges and three quantum point contacts. In the absence of QPC3 the set-up Fig. 4 is equivalent to Fig. 3. If weak tunneling at QPC3 is allowed the system exhibits strong telegraphic noise which carries information about fractional statistics. In contrast to Ref. 
\onlinecite{kane} we investigate the set-up with two point contacts that was studied experimentally \cite{MZ}. In our case not only the noise but also the average current is strongly sensitive to fractional statistics.

\section{Model}

\subsection{Effective action for electron tunneling case}

We first consider the system represented by Fig. 1a). Its low-energy behavior can be described by the chiral Luttinger liquid model \cite{wen}. Fig. 5 illustrates the model.  The two edges 1 and 2 correspond to two chiral Luttinger liquids with the same propagation direction. The dashed lines describe quantum point contacts
where electrons tunnel between the edges. Note that the respective distances $L$ and $L+a$ between the point contacts along the two edges  are different. The Lagrangian assumes the standard form \cite{wen}

\begin{equation}
\label{1}
L=-\frac{\hbar}{4\pi}\int dx dt \sum_{k=1,2} [\partial_t\phi_k\partial_x\phi_k+v(\partial_x\phi_k)^2]-\int dt (T_1+T_2),
\end{equation}
where $T_{1}$ and $T_2$ are tunneling operators, $v$ is the excitation velocity along the edges, and $\{\phi_k\}$ represent two chiral Bose fields which satisfy the following commutation relations

\begin{equation}
\label{2}
[\phi_l(x_l, t=0),\phi_p(x_p,t=0)]=i\pi\delta_{lp}{\rm sign}(x_l-x_p).
\end{equation}
The Bose fields are related to the charge densities $\rho_l$ through

\begin{equation}
\label{3}
\rho_l=(\sqrt{\nu_l}e/2\pi)\partial_x\phi_l,
\end{equation}
where 
$e$ is an electron charge and
$\nu_l=1/(2m_l+1)$ are the filling factors
of the QHE puddle defined by edge 2e and the QHE strip bounded by edge 1e.
The tunneling operators \cite{wen}

\begin{eqnarray}
\label{4}
T_1=\Gamma_1\exp(i[\phi_1(0,t)/\sqrt{\nu_1}-\phi_2(0,t)/\sqrt{\nu_2}])+h.c.; &&  \nonumber\\
T_2=\Gamma_2\exp(i[\phi_1(L,t)/\sqrt{\nu_1}-\phi_2(L+a,t)/\sqrt{\nu_2}])+h.c
\end{eqnarray}
are proportional to the electron annihilation and creation operators $\sim\exp(\pm i\phi_l/\sqrt{\nu_l})$.
The action also includes operators describing simultaneous tunneling of several electrons
but they do not play a significant role 
at low energies.

In the presence of a magnetic flux $\Phi$ through the region A-QPC2-B-QPC1-A, the tunneling amplitude
$\Gamma_2$ should be multiplied by the phase factor $\exp(2\pi i\Phi/\Phi_0)$, where $\Phi_0=hc/e$ is the flux quantum
\cite{footnote}. This phase factor describes the difference of the phases $e/(\hbar c)\int_{{\rm QPC1}}^{\rm QPC2} \vec A d\vec l$ accumulated by the electrons moving along two edges between the point contacts. For one edge the integration path is
QPC1-A-QPC2, and for the other edge the integration path is QPC1-B-QPC2 (Fig. 1).

The voltage bias $V$ results in the chemical potential difference $\mu_1-\mu_2=eV$ between edges 1e and 2e. 
We would like to emphasize that the potential difference between the edges
is determined by the voltage difference between the sources
S1 and S2 in both set-ups Fig. 1a) and Fig. 1b); however, the shape of the edges 1e and 2e, Fig. 1a), is different from the shape of the edges 1q and 2q, Fig. 1b).
We will use the interaction representation which makes both chemical potentials equal and introduces time-dependence into the tunneling amplitudes: 

\begin{equation}
\label{5}
\Gamma_1,\Gamma_2\sim\exp(-\frac{ieVt}{\hbar}).
\end{equation}
The difference of the chemical potentials results also in the difference of the average densities between the edges.
We assume that edge 2 is connected to the ground. Then the average density $\rho_2\sim\langle\partial_x\phi_2\rangle$ is zero in the limit of weak tunneling. It will be convienient for us to shift the Bose field on the first edge
$\phi_1(x)\rightarrow\phi_1(x)-f(x)$ by a function $f(x)$ of the coordinate in such a way that $\langle\partial_x\phi_1\rangle$ becomes zero.
The mean charge density on edge 1 is proportional to the voltage bias and can be found from the minimization of the Hamiltonian

\begin{equation}
\label{z1}
H=\hbar\int dx\left[\frac{v}{4\pi}(\partial_x\phi_1)^2-\frac{eV\sqrt{\nu_1}}{2\pi\hbar}\partial_x\phi_1\right].  
\end{equation}
This yields
$q=\langle\partial_x\phi_1\rangle=eV\sqrt{\nu_1}/(\hbar v)$. We next
 shift the field $\phi_1\rightarrow\phi_1-qx$ such that $\langle\partial_x\phi_1\rangle$ vanishes.
At the same time $\Gamma_2$ is multiplied by the factor $\exp(ieVL/\hbar v)$.

Hereafter we assume that $\nu_1\ge\nu_2$. The case $\nu_1=\nu_2=1$ corresponds to the free electron problem \cite{FMZ1,FMZ2}.

\subsection{Effective action for quasiparticle tunneling case}

We now consider the system depicted in Fig. 1b). The model is given by the chiral Luttinger liquid action (\ref{1})
with a different choice of the tunneling operators \cite{wen}. Several modifications immediately follow from the fact that the quasiparticle charge $\nu e$ differs from the electron charge. 1) The tunneling operators should be expressed via the quasiparticle annihilation and creation operators $\sim\exp(\pm i\sqrt{\nu}\phi_l)$; 2) the flux-dependent phase factor in
$\Gamma_2$ is now $\exp(2\pi\nu i \Phi/\Phi_0)$; 3) the phase factor 
$\exp(ie\nu VL/\hbar v)$ should be used instead of $\exp(ieVL/\hbar v)$;
and 4) in the interaction representation the time-dependence of the tunneling amplitudes becomes

\begin{equation}
\label{6}
\Gamma_1,\Gamma_2\sim\exp(-\frac{ie\nu Vt}{\hbar}).
\end{equation}
The fifth difference from Eqs. (\ref{4}) consists in the introduction of two Klein factors $\kappa_1$ and $\kappa_2$
(a related model without Klein factors has been considered in Ref. \onlinecite{tm}):

\begin{eqnarray}
\label{7}
T_1^q=\Gamma_1\kappa_1\exp(i\sqrt{\nu}[\phi_1(0,t)-\phi_2(0,t)])+h.c.; &&  \nonumber\\
T_2^q=\Gamma_2\kappa_2\exp(i\sqrt{\nu}[\phi_1(L,t)-\phi_2(L+a,t)])+h.c,
\end{eqnarray}
where the commutation relations are 

\begin{eqnarray}
\label{8}
\kappa_1\kappa_2=\exp(-2\pi\nu i)\kappa_2\kappa_1; && \nonumber\\
\kappa_1\kappa_2^+=\kappa_2^+\kappa_1\exp(2\pi\nu i).
\end{eqnarray}

These commutation relations can be understood from a locality argument similar to Ref. \onlinecite{kane}. Two tunneling operators affect two distant parts of the system. Hence, for any reasonable model 
$[T_1^q,T_2^q]=0$. 
One can calculate
$[T_1^q, T_2^q]$ employing the commutation relations for the Klein factors, the commutation relations 
Eq. (\ref{2}) for the Bose fields and the Baker-Hausdorff formula. This results in $[T_1^q, T_2^q]=[T_1^q,(T_2^q)^+]=0$ 
provided that Eq. (\ref{8}) is satisfied.
A different formulation of the same argument can be found in Ref. \onlinecite{kane}. 
Ref. \onlinecite{pa} discusses how Klein factors which ensure commutativity of tunneling operators can be derived using duality between weak quasiparticle tunneling and strong electron tunneling. 

The Klein factors serve as a manifestation of fractional statistics and are absent in the case of fermion tunneling, section III.A.
The importance of Klein factors in quantum Hall systems with more than two edges has been emphasized previously 
\cite{HBT,kane,pa,nfll,klein}. In our problem, the Klein factors are necessary even though there are only two edges. Note 
that in the 1a) setup no Klein factors are needed. 
Indeed, the operators $T_1$ and $T_2$ defined by Eq. (\ref{4}) commute.
Commutativity of the quasiparticle tunneling operators is also ensured without Klein factors for the standard 
Aharonov-Bohm interferometer geometry \cite{cfksw}, Fig. 3.

In our calculations we will use the following representation of the Klein factors by $1/\nu\times 1/\nu$ matrices:

\begin{equation}
\label{s3e1}
\kappa_1= 
\left( \begin{array}{cccccc}
0 & 1 & 0 & 0 &\dots & 0 \\
0 & 0 & 1 & 0 & \dots & 0 \\
\dots & \dots & \dots & \dots & \dots &\dots \\
0 & \dots &\dots & 0 & 1 & 0 \\
0 & \dots & \dots & \dots & 0 & 1 \\ 
1 & 0 & \dots & \dots & \dots & 0\end{array} \right);~~~~~~~
\kappa_2=
\left( \begin{array}{cccccc}
0 & \psi & 0 & 0 &\dots & 0 \\
0 & 0 & \psi^2 & 0 & \dots & 0 \\
\dots & \dots & \dots & \dots & \dots & \dots \\
0 & \dots &\dots & 0 & \psi^{1/\nu-2} & 0 \\
0 & \dots & \dots & \dots & 0 & \psi^{1/\nu-1} \\ 
1 & 0 & \dots & \dots & \dots & 0\end{array} \right),
\end{equation}
where $\psi=\exp(-2\pi\nu i)$. One can easily check that the above matrices satisfy the commutation relations (\ref{8}).
The tunneling operators (\ref{7}) with the Klein factors (\ref{s3e1}) can be understood as products of a quasiparticle creation operator, an annihilation operator and a phase factor which includes the statistical phase accumulated
during the tunneling event.

The charge distribution on the edges is determined by the fields $\{\phi_l(x)\}$, the total charge being determined by the zero modes of these fields. 
In the set-up depicted in Fig. 1a) 
the charge distribution completely describes all states in the Hilbert space on which the effective low-energy Hamiltonian acts. 
On the other hand, our discussion in section II shows that in the case of quasiparticle tunneling one needs to 
specify both the charge distribution and the effective statistical flux through the interferometer.
Hence, the corresponding Hilbert space is the product of the space $V_{\rm charge}$ on which the 
Bose operators $\phi_l$ act
and the space $V_{\rm flux}$ on which the Klein factors act. 
As is clear from the size of the matrices (\ref{s3e1}) the dimensionality of the latter space is 
${\rm dim}~V_{\rm flux}=1/\nu$. This agrees with our discussion 
in section II where we found that the interferometer has $1/\nu$ classes of states characterized by different probabilities of quasiparticle tunneling. 

Calculations based on the model (\ref{1}) with the tunneling operators (\ref{7}) confirm Eq. (\ref{s2e7})
of Section II. The Klein factors keep track of fractional statistics and are crucial for this result. In the absence of the Klein factors, i.e. for fractionally charged particles which do not obey fractional statistics, one gets qualitatively different results \cite{tm}.





\section{Electron tunneling}

We now study the electric current between the two edges. 
We consider the geometry depicted in Fig. 1a).
The current operator

\begin{eqnarray}
\label{10}
\hat I={\frac{d}{dt}{\hat Q_1}}=\frac{i}{\hbar}[\hat H,\hat Q_1]= 
\frac{i e \Gamma_1}{\hbar}\exp(-ieVt/\hbar)\exp[i(\phi_1(0)/\sqrt{\nu_1}-\phi_2(0)/\sqrt{\nu_2})] & & \nonumber\\ 
+
\frac{i e \Gamma_2}{\hbar}\exp(ieVL/\hbar v+2\pi i\Phi/\Phi_0)\exp(-ieVt/\hbar)\exp[i(\phi_1(L)/\sqrt{\nu_1}-\phi_2(L+a)/\sqrt{\nu_2})]+ h.c.,
\end{eqnarray}
where $Q_1$ is the total charge of the first edge and $H$ the Hamiltonian.
For our non-equilibrium problem we employ the Keldysh technique \cite{Keldysh}.
To this end we assume that the tunneling amplitudes $\Gamma_{1,2}=0$ at the moment of time $t=-\infty$, and are subsequently turned on gradually. At $t=-\infty$ the system is in thermal equilibrium at temperature $k_B T$ and chemical potential difference $eV$ between the edges. The initial equilibrium state determines the bare Keldysh Green functions
which will be used in the perturbative calculations below.

The current at $t=0$ is

\begin{equation}
\label{11}
I={\rm Tr} [\hat\rho S(-\infty,0)\hat I S(0,-\infty)],
\end{equation}
where $\hat\rho$ is the initial density matrix and $S(0,-\infty)={\rm T}\exp(- 
i\int \hat H dt/\hbar)$ the evolution operator.
Expanding the latter to first order in the tunneling amplitudes one finds

\begin{eqnarray}
\label{12}
I=\frac{e}{\hbar^2}\int_{-\infty}^{+\infty}dt(|\Gamma_1|^2+|\Gamma_2|^2)\exp(-\frac{ieVt}{\hbar})[F(0,0,t)-F(0,0,-t)] & & \nonumber \\ 
+\frac{e}{\hbar^2}\int_{-\infty}^{+\infty}dt\left\{
\Gamma_1\Gamma_2^*\exp(-\frac{ieVL}{\hbar v}-2\pi i\Phi/\Phi_0)
\exp(-\frac{ieVt}{\hbar})[F(-L,-(L+a),t)-F(L,L+a,-t)]
+ h.c. \right\},
\end{eqnarray}
where 

\begin{eqnarray}
\label{13}
F(b,c,t)={\rm Tr}[\hat\rho \exp(i\phi_1(x=b,t)/\sqrt{\nu_1})\exp(-i\phi_1(x=0,0)/\sqrt{\nu_1})]
& & \nonumber \\
\times {\rm Tr}[\hat\rho \exp(i\phi_2(x=c,t)/\sqrt{\nu_2})\exp(-i\phi_2(x=0,0)/\sqrt{\nu_2})].
\end{eqnarray}

Let us first consider the zero temperature case. The correlation function is given by \cite{bos} 

\begin{equation}
\label{14}
F(b,c,t)=\frac{\tau_c^{1/\nu_1}}{[\delta+i(t-b/v)]^{1/\nu_1}}\frac{\tau_c^{1/\nu_2}}{[\delta+i(t-c/v)]^{1/\nu_2}},
\end{equation}
where $\delta$ is an infinitesimal positive constant, and $\tau_c$ is the ultra-violet cut-off \cite{foot2}.
With the above expression we find

\begin{equation}
\label{15}
I=I_0+I_\Phi,
\end{equation}
where

\begin{eqnarray}
\label{15a}
I_0=-\frac{2\pi e\tau_c}{\hbar^2}(|\Gamma_1|^2+|\Gamma_2|^2)\frac{[\tau_c eV/\hbar]^{1/\nu_1+1/\nu_2-1}}{(1/\nu_1+1/\nu_2-1)!},
\end{eqnarray}

\begin{eqnarray}
\label{15b}
I_\Phi= \frac{2\pi ie\tau_c^{1/\nu_1+1/\nu_2}}{\hbar^2} (-1)^{1/[2\nu_1]+1/[2\nu_2]}\times& & \nonumber \\
\times\left\{
\Gamma_1\Gamma_2^*\exp(-2\pi i \Phi/\Phi_0)
\left[
\frac{1}{(1/\nu_1-1)!}\frac{d^{1/\nu_1-1}}{dz^{1/\nu_1-1}}|_{z=0}\frac{\exp(-\frac{ieVz}{\hbar})}{(z+a/v)^{1/\nu_2}}
+\frac{1}{(1/\nu_2-1)!}\frac{d^{1/\nu_2-1}}{dz^{1/\nu_2-1}}|_{z=-a/v}\frac{\exp(-\frac{ieVz}{\hbar})}{z^{1/\nu_1}}
\right]
-h.c.
\right\}
\end{eqnarray}
The current oscillates as function of the magnetic flux with period $\Phi_0$. At low voltages it follows the power law

\begin{equation}
\label{s4e1}
I\sim V^{1/\nu_1+1/\nu_2-1} 
\end{equation}
(see Appendix A; cf. Ref. \onlinecite{geller}). Fig. 6 illustrates the $I-V$ curves for $\nu_1=\nu_2=1$, $\nu_1=\nu_2=1/3$ and 
$\nu_1=1,\nu_2=1/3$. In the case when $\nu_1=\nu_2$, similar expressions for other interferometer geometries were obtained
in Refs. \onlinecite{cfksw,geller}. One can easily see that Eq. (\ref{s2e1}) 
follows from Eqs. (\ref{15}-\ref{15b}).

Notice that only the difference $a$ of the lengths of the edges enters the above expression for the current while the total length $L$ of the first edge drops out. In the standard geometry of an Aharonov-Bohm interferometer \cite{cfksw} the flux-dependent contribution to the current depends on the total interferometer size and decreases with the system's size. Thus, quantum interference effects cannot be observed at large  system's sizes. In the Mach-Zehnder geometry, quantum interference 
can be observed for $L\gg a$ since only the difference of the phases accumulated by the particles moving along two edges is important.

The case of non-zero temperature is considered in Appendix B. The flux-dependent ``interference'' contribution $I_\Phi$ to the current vanishes at large $ak_B T/(hv)$. In the opposite limit of $ak_B T/(hv)\ll 1$ as well as for
$ak_B T\sim hv$, the flux-independent contribution $I_0$ and $I_\Phi$ are related by Eq. (\ref{s2e1}). 
The linear conductance at low temperatures and low voltages $eV\ll k_B T$ scales as

\begin{equation}
\label{16}
G={I}/{V}\sim [k_B T]^{1/\nu_1+1/\nu_2-2}.
\end{equation}

\section{Quasiparticle tunneling}

We now consider the geometry depicted in Fig. 1b).
The current through the interferometer oscillates as function of the magnetic flux.
In the first subsection below we determine the oscillation period. It turns out to be the same as in the case of electron 
tunneling and equals one flux quantum $\Phi_0$. Next, we use the perturbation theory to
calculate the current as  function of the voltage, temperature and the distance between the quantum point contacts. 
We confirm Eq. (\ref{s2e14}). The dependence of the current (\ref{s2e14}) on the tunneling amplitudes $\Gamma_1$ and 
$\Gamma_2$ is nonanalytic. Such dependence cannot be obtained in any finite order of the perturbation theory. Thus,
we have to sum up an infinite set of diagrams. In subsection V.B this is made for the simplest case of zero temperature
and low voltage, $eVa\ll\hbar v$, at the filling factor $\nu=1/3$. The general case is considered in the final 
subsection.

\subsection{Period of Aharonov-Bohm oscillations}

The tunneling current operator can be found with the same method as in the previous section:

\begin{eqnarray}
\label{17}
\hat I={\frac{d}{dt}{\hat Q_1}}=\frac{i}{\hbar}[\hat H,\hat Q_1]= 
\frac{ie\kappa_1\nu\Gamma_1}{\hbar}\exp(-ie\nu Vt/\hbar)\exp[i\sqrt\nu(\phi_1(0)-\phi_2(0))] & & \nonumber\\ 
+
\frac{ie\kappa_2\nu\Gamma_2}{\hbar}\exp(\frac{ie\nu VL}{\hbar v}+2\pi\nu i\Phi/\Phi_0)\exp(-ie\nu Vt/\hbar)\exp[i\sqrt\nu(\phi_1(L)-\phi_2(L+a))]+ h.c.
\end{eqnarray}

The average current is given by Eq. (\ref{11}). Since the tunneling amplitudes $\Gamma_l$ are small, we will 
employ perturbation theory. To lowest (second) nonzero order in $\Gamma_l$ the resulting contributions are 
proportional to $|\Gamma_1|^2$ and $|\Gamma_2|^2$. The cross-terms proportional to $\Gamma_1^*\Gamma_2$
and $\Gamma_1\Gamma_2^*$ vanish. Indeed, each term of the perturbative expansion is proportional to the product of the average of some function of the Bose-operators $\{\phi_l\}$ and the average of some product of Klein factors. The averages are determined by the initial density matrix $\hat\rho$, Eq. (\ref{11}). The latter depends on the 
effective action Eq. (\ref{1}) at $t=-\infty$. That action does not contain Klein factors. Hence, the density matrix 
$\hat\rho=\hat\rho_\phi\hat\rho_\kappa$, where $\hat\rho_\phi$ and $\hat\rho_\kappa$ act on the spaces 
$V_{\rm charge}$ and $V_{\rm flux}$ respectively, and $\hat\rho_\kappa$ 
is proportional to the unit matrix
at any finite temperature. 
Thus, the cross-terms are proportional to the expressions of the form ${\rm Tr}[\kappa_l\kappa_m^+]$, where $l\ne m$. Such traces are zero. 
This is readily seen from the following argument.
For any two linear operators 
${\rm Tr} AB ={\rm Tr} BA$. Hence, 
${\rm Tr} [ \kappa_1\kappa_2^+ ]={\rm Tr}[ \kappa_2^+\kappa_1]$. At the same time it follows 
from Eq. (\ref{8}) that 
${\rm Tr} [ \kappa_1\kappa_2^+ ]=\exp(2\pi\nu i){\rm Tr}[ \kappa_2^+\kappa_1 ]$. Hence,
${\rm Tr} [ \kappa_1\kappa_2^+ ]=0$. 
It follows that there are no cross-terms in the second order perturbation theory.

The above result implies that the second-order terms are independent of the magnetic flux. 
%
%
The flux dependence of the current emerges only in higher orders of the perturbation theory. Each term of the perturbative expansion of Eq. (\ref{11}) contains a product of Klein factors. Let the number of the Klein factors
$\kappa_l^+$ in a given term $I_\alpha$ be $n_l^+$ and the number of the factors $\kappa_l$ be $n_l^-$ . 
Each of the four numbers $n_l^{\pm}$ indicates the power in which the respective coefficient $\Gamma_1$, $\Gamma_2$, $\Gamma_1^*$ or $\Gamma_2^*$ enters in $I_\alpha$. Nonzero terms of the perturbation series describe the processes which do not change the edge charges. 
Hence, $n_1^+ - n_1^-=-(n_2^+ - n_2^-)$. 

We next show that
$\nu (n_1^+ - n_1^-)^2$ is an integer for any non-vanishing $I_\alpha$. Consider the trace $W$ of the product of the Klein factors in the term $I_\alpha$.
We can move all operators $\kappa_1^{+},\kappa_1$ to the left of all operators $\kappa_2^{+},\kappa_2$. 
This will produce a phase factor $\exp(i\gamma)$. The trace can be represented as $W=\exp(i\gamma){\rm Tr} K_1 K_2$, where $K_1$ is a product of $\kappa_1^{+},\kappa_1$ and $K_2$ is a product of $\kappa_2^{+},\kappa_2$. We know that 

\begin{equation}
\label{20}
{\rm Tr} K_1 K_2={\rm Tr}K_2 K_1. 
\end{equation}
At the same time, one can move all Klein factors $\kappa_2^{+},\kappa_2$ in the product 
$K_1 K_2$ to the left of the operator $K_1$ using the commutation relations Eq. (\ref{8}). This yields 

\begin{equation}
\label{21}
K_1 K_2=\exp[2\pi\nu i (n_1^+ - n_1^-)^2] K_2 K_1.
\end{equation}
Eqs. (\ref{20},\ref{21}) show that $I_\alpha \sim {\rm Tr} K_1 K_2\ne 0$ 
only if $\nu (n_1^+ - n_1^-)^2$ is an integer.

If $\nu=1/[{\rm odd~prime~number}]$, including experimentally relevant $\nu=1/3$ and $\nu=1/5$, 
the above result means that 
$(n_1^+-n_1^-)=-(n_2^+ -n_2^-)=q/\nu$, where $q$ is an integer.
The term $I_\alpha\sim \Gamma_2^{n_2^-}(\Gamma_2^*)^{n_2^+}\sim \exp(2\pi\nu i\Phi/\Phi_0[n_2^- - n_2^+])$.
Hence, it follows that the current is a periodic function of the 
magnetic flux with period $\Phi_0$. As shown in the following subsections the period is the same for any 
$\nu=1/(2m+1)$. This can also be verified by a direct calculation of the trace (\ref{20}) using (\ref{s3e1}).
Such periodicity agrees with the Byers-Yang theorem \cite{BY,footnote2}. If fractionally charged
quasiparticles did not obey fractional statistics, i.e. if there were no Klein factors, the period would be
$\Phi'=\Phi_0/\nu$ (cf. Ref. \onlinecite{tm}).

Our effective hydrodynamic action (\ref{1},\ref{7}) is applicable only for low
temperatures and voltages. We are going to use the perturbation theory in $\Gamma_{1,2}$. It turns out that the lowest order contribution to the current scales as $I\sim V\Gamma_{1,2}^2{\rm max}(eV,k_B T)^{\alpha}$, where $\alpha=2\nu-2$ is negative. The perturbation theory can be used for the calculation of the tunneling current only when the tunneling current is much smaller than the current $\nu e^2 V/h$ incoming from the sources. 
Thus, our calculations are valid provided that 
$\Gamma_{1,2}{\rm max}(eV,k_B T)^{\nu-1}\ll \left(\frac{\hbar}{\tau_c}\right)^\nu$.

We will see that interference effects can be observed only for small enough $a\sim hv/{\rm max}(eV,k_B T)$. 
This condition is similar to the restriction on the total interferometer size in Ref. \onlinecite{cfksw}.
In our case the restriction on the total size $L$ is weaker. As is clear from section II our analysis is based on the assumption that the time $\Delta t$ between tunneling events exceeds the time $L/v$ needed to a quasiparticle to travel from QPC1 to QPC2. Thus, 
$L<v\Delta t\sim \frac{v\hbar^2}{\Gamma_{1,2}^2\tau_c} \left(\frac{\hbar}{{\rm max}(eV, k_B T)\tau_c}\right)^{2\nu-1}$

\subsection{${\bf \nu=1/3}$, ${\bf {\it T}=0}$, ${\bf\it eVa\ll \hbar v }$}

We will use the expansion of the current, Eq. (\ref{11}), in powers of the tunneling amplitudes 
$\Gamma_{1,2}$. Only contributions proportional to even powers of the tunneling operators are non-zero. One might 
naively expect on the basis of power counting that the terms of order $2n$ in $\Gamma_{1,2}$ scale as 
$V|\Gamma_1|^p |\Gamma_2|^{2n-p}\times[{\rm max}(eV, T)]^{2n(\nu-1)}$,
if $L\sim vh/{\rm max}(eV, k_B T)$. 
This is however not the case beyond the second perturbative order. In fact, the 4th order contribution (as well as higher orders) is infinite. 
To demonstrate this we use an analogy between the power expansion of the average current and the partition function of a Coulomb gas \cite{CQ}. Positively and negatively charged ``particles'' correspond to the
tunneling events from edge 1q to edge 2q and from edge 2q to edge 1q respectively. The coordinates of the particles
 correspond to the times of the tunneling
events. The charges can be located on both branches of the Keldysh contour, Fig. 7. The particles on the top and lower
branches
emerge from the expansions of $S(0,-\infty)$ and $S(-\infty,0)$ in Eq. (\ref{11}) respectively.
 As discussed in Ref. \onlinecite{CQ}, attraction between oppositely charged particles
 binds them in pairs. 
The typical pair size is of order $\hbar/{\rm max}(eV,k_B T)$ at low temperatures and voltages \cite{footnote3}
(if $L>hv/{\rm max}(eV, k_B T)$ then pairs of size $L/v$ are possible). 
When 
$\frac{\Gamma_{1,2}\tau_c}{\hbar}\left [\frac{{\rm max} (eV,k_B T) \tau_c}{\hbar} \right ]^{\nu-1}\ll 1$
and
$L\ll \frac{v\hbar^2}{\Gamma_{1,2}^2\tau_c} \left(\frac{\hbar}{{\rm max}(eV, k_B T)\tau_c}\right)^{2\nu-1}$, 
the pairs
are dilute and do not overlap. Hence, any term in the perturbation expansion reduces to 
the trace of a product of Klein factors times a product of the two-point
correlation functions 

\begin{eqnarray}
\label{s5e1}
F[b=x_1-x_2;c=(x_1-x_2)(L+a)/L;t=t_1-t_2]=
{\rm Tr}\{\hat\rho \exp[i\sqrt{\nu}\phi_1(x=x_1,t_1)]\exp[-i\sqrt{\nu}\phi_1(x=x_2,t_2)]\}\times
& & \nonumber \\
{\rm Tr}\{\hat\rho \exp[i\sqrt{\nu}\phi_2(x=x_1(L+a)/L,t_1)]\exp[-i\sqrt{\nu}\phi_2(x=x_2(L+a)/L,t_2)]\}
\end{eqnarray}
corresponding to the bound pairs. This product should be integrated over the times of all tunneling events, i.e.
the positions of the charges. This integration can be separated into the product of
the integrals over the dipole sizes $(t_1-t_2)$,
Eq. (\ref{s5e1}), and the integrals over the dipole positions $(t_1+t_2)/2$.

Fig. 7 illustrates the 4th order contribution to the current. One charge is located at $t=0$ and corresponds to the current operator in Eq. (\ref{11}). It forms a dipole with an opposite charge at the point 
$\tau_0\sim\hbar/{\rm max}(eV,k_B T)$.
Two more mutually opposite charges are located at the points $t_1\pm\tau_1/2$, where $\tau_1$ is the dipole size.
The charges can reside on the same or different branches of the Keldysh contour as shown in Figs. 7a), 7b) and 7c).
The integral over $t_1$ diverges. In the absence of the Klein factors the infinite integrals corresponding to the configurations
Fig. 7a), 7b) and 7c) cancel but this is no longer the case when the Klein factors are included. Hence, the 4th order
contribution is infinite. The same argument applies to higher-order contributions. 
Certainly, the sum of all perturbative orders must be finite but as is clear from section II it is not an analytic function of $\Gamma_{1,2}$. Below we develop a method to sum up all orders of the perturbation theory.

We begin with the simplest case when ${ \nu=1/3}$; ${ { T}=0}$; ${ eVa,eVL\ll \hbar v }$. The same approach will 
be applied to
the general situation in section V.C. The first condition simplifies the structure of the Klein factors (\ref{s3e1}) 
which become $3\times 3$ matrices at filling factor $1/3$. The second condition allows us to use a simpler zero-temperature expression for the correlation function

\begin{eqnarray}
\label{s5efg1}
F(b,c,t_1-t_2)={\rm Tr}[\hat\rho \exp(i\sqrt{\nu}\phi_1(x=b,t_1))\exp(-i\sqrt{\nu}\phi_1(x=0,t_2))]\times
& & \nonumber \\
{\rm Tr}[\hat\rho \exp(i\sqrt{\nu}\phi_2(x=c,t_1))\exp(-i\sqrt{\nu}\phi_2(x=0,t_2))]=
\frac{\tau_c^{\nu}}{[\delta+i(t_1-t_2-b/v)]^{\nu}}\frac{\tau_c^{\nu}}{[\delta+i(t_1-t_2-c/v)]^{\nu}}
\end{eqnarray}
The third condition makes it possible to neglect the distances $L$ and $(L+a)$ between the point contacts. 
Indeed, the tunneling operator $T_2^q=:\Gamma_2\kappa_2\exp(i\sqrt{\nu}[\phi_1(L,t)-\phi_2(L+a,t)]):+h.c$,
Eq. (\ref{7}), can be rewritten
as

\begin{eqnarray}
\label{s5e2}
T_2^q=:\Gamma_2\kappa_2\exp\left(i\sqrt{\nu}[\phi_1(0,t)-\phi_2(0,t)]+i\sqrt{\nu}
\sum_{k=1}^\infty{[\frac{\partial^k\phi_1}{\partial x^k}L^k-\frac{\partial^k\phi_2}{\partial x^k}(L+a)^k]}/{k!}\right):   +h.c.= & & \nonumber \\
:\Gamma_2\kappa_2\exp(i\sqrt{\nu}[\phi_1(0,t)-\phi_2(0,t)])
\left\{1+i\sqrt{\nu}\left[\frac{\partial\phi_1}{\partial x}L-
\frac{\partial\phi_2}{\partial x}(L+a)\right]
+\dots\right\}:
+h.c.,
\end{eqnarray}
where the ellipses denote higher order gradients of $\phi_l$
and the colons denote normal ordering. 
After the substitution of Eq. (\ref{s5e2})
in the perturbative expansion of the current one can compare the contributions from the terms containing
derivatives of $\phi_l$ with the contributions from the terms which do not contain such derivatives.
Power counting shows that the terms with derivatives are suppressed by the factors of order $(LeV/\hbar v)^k;
([L+a]eV/\hbar v)^k$ and hence can be neglected. This conclusion agrees with the results of section V.C for arbitrary
voltages, temperatures and interferometer sizes.

Thus, we can use an effective single impurity model. The tunneling operator 
$O  = (T_1^q+T_2^q)$, Eqs. (\ref{7},\ref{s3e1}),
assumes the form

\begin{equation}
\label{s5e3}
O=O_-+O_+;~~O_- =  O_+^\dagger =\kappa_-A_-,
\end{equation}
where

\begin{equation}
\label{s5e4}
A_-=A_+^\dagger=\exp(i\sqrt{\nu}[\phi_1(0,t)-\phi_2(0,t)])
\end{equation}
and the (non-unitary) operator $\kappa_-$ is

\begin{equation}
\label{s5e5}
\kappa_-=\kappa_+^\dagger=\Gamma_1\kappa_1+\Gamma_2\exp(2\pi\nu i \Phi/\Phi_0)\kappa_2=
\left( \begin{array}{ccc}
0 & C_1 & 0  \\
0 & 0 & C_2  \\ 
C_3 & 0 & 0 \end{array} \right)
\end{equation}
with $C_1=\Gamma_1+\Gamma_2\exp(2\pi\nu i \Phi/\Phi_0)\psi$, $C_2=\Gamma_1+\Gamma_2\exp(2\pi\nu i \Phi/\Phi_0)\psi^2$,
$C_3=\Gamma_1+\Gamma_2\exp(2\pi\nu i \Phi/\Phi_0)$ and $\psi=\exp(-2\pi i/3)$. 
In what follows we will denote the basis vectors
$(1,~0,~0)$, $(0,~1,~0)$ and $(0,~0,~1)$ as
 $\langle 1|$, $\langle 2|$ and $\langle 3|$ respectively.
Since we neglect $L$ and $(L+a)$
in the rest of this section, we can use the correlation function (\ref{s5efg1}) in the simplest limit $b,c=0$.

The contribution $I_{2N}$ of order $2N$ to the current corresponds to the charge configuration with $N$ dipoles. One dipole of size $\tau_0$ is located at $t=0$.
The remaining dipoles of sizes $\tau_1,\dots,\tau_{N-1}$ are located at 
$0>t_1>t_2>\dots>t_{N-1}$. One finds

\begin{eqnarray}
\label{s5e6}
I_{2N}=-\sum_{b_0=\pm 1}\sum_{b^\pm_k=\pm 1}\sum_{\sigma[0]=\pm}
\sigma[0]\frac{\nu e i^{2N}}{\hbar^{2N}}
\int_{-\infty}^0 dt_1 \int_{-\infty}^{t_1} dt_2 \dots 
\int_{-\infty}^{t_{N-2}} dt_{N-1} \int_{-\infty}^0 d\tau_0
\int_{-\infty}^{+\infty} d\tau_1\dots d\tau_{N-1} 
b_0\Pi_k b_k^+b_k^-
& & \nonumber\\
{\rm Tr} 
\left [
\hat\rho {\rm T}_c 
O_{\sigma[0]}(t=0) O_{-\sigma[0]}(\tau_0;b_0)
O_-(t_1+\frac{\tau_1}{2};b^-_1)O_+(t_1-\frac{\tau_1}{2};b^+_1)\dots
O_-(t_{N-1}+\frac{\tau_{N-1}}{2};b^-_{N-1})
O_+(t_{N-1}-\frac{\tau_{N-1}}{2};b^+_{N-1})
\right ],
\end{eqnarray}
where ${\rm T}_c$ denotes time-ordering along the Keldysh contour, 
$b_0,b^+_k,b^-_k=+1$ 
correspond to the bottom branch of the Keldysh contour
and $b_0,b^+_k,b^-_k=-1$ correspond to the top branch.
 
The trace in Eq. (\ref{s5e6}) can be factorized as 

\begin{equation}
\label{s5e0}
{\rm Tr}[...]=\Pi_A\Pi_\kappa, 
\end{equation}
where
$\Pi_A$ stays for the trace of a product of operators $A_{\pm}$, and 
$\Pi_\kappa$ denotes the trace of a time-ordered product of the Klein factors.
The former trace can be further factorized into a product of the two-point correlation functions (\ref{s5efg1}) corresponding to each dipole. We next simplify the expression for $\Pi_\kappa$,

\begin{equation}
\label{s5e7}
\Pi_\kappa={\rm Tr}\left [\hat\rho_\kappa {\rm T}_c
\kappa_{\sigma[0]}(t=0) \kappa_{-\sigma[0]}(\tau_0;b_0)
\kappa_-(t_1+\frac{\tau_1}{2};b^-_1)\kappa_+(t_1-\frac{\tau_1}{2};b^+_1)
\dots
\kappa_-(t_{N-1}+\frac{\tau_{N-1}}{2};b^-_{N-1})
\kappa_+(t_{N-1}-\frac{\tau_{N-1}}{2};b^+_{N-1})
\right ].
\end{equation} 
Using the matrix elements of the Klein factors the above equation can be rewritten as

\begin{equation}
\label{s5e8}
\Pi_\kappa=\frac{1}{3}\sum_{e_N=1,2,3}\sum_{e_k^{t/b}=1,2,3;k=1,\dots,N-1}
I(e_1^t,e_1^b) B(e_1^t,e_1^b;e_2^t,e_2^b)\dots 
B(e_{N-2}^t,e_{N-2}^b;e_{N-1}^t,e_{N-1}^b)
B(e_{N-1}^t,e_{N-1}^b;e_{N}^t=e_N,e_{N}^b=e_N),
\end{equation} 
where the factor $1/3$ comes from the condition ${\rm Tr}\hat\rho_\kappa=1$;
$I(e_1^t,e_1^b)=
\langle e_1^b| 
T_c \kappa_{\sigma[0]}(t=0) \kappa_{-\sigma[0]}(\tau_0;b_0)
| e_1^t\rangle$ and

\begin{eqnarray}
\label{s5e9}
B(e_{k}^t,e_{k}^b;e_{k+1}^t,e_{k+1}^b)=
\langle e_{k+1}^b| e_k^b\rangle 
\langle e_k^t| T_c \kappa_-(t_k+\tau_k/2;-1)\kappa_+(t_k-\tau_k/2;-1) |e_{k+1}^t \rangle 
{\rm ~~ for ~~} b_k^+=b_k^-=-1 & & \nonumber\\
B(e_{k}^t,e_{k}^b;e_{k+1}^t,e_{k+1}^b)=
\langle e_{k+1}^b| T_c \kappa_-(t_k+\tau_k/2;+1)\kappa_+(t_k-\tau_k/2;+1) |  e_k^b\rangle 
\langle e_k^t| e_{k+1}^t\rangle {\rm ~~ for ~~} b_k^+=b_k^-=+1  & & \nonumber\\
B(e_{k}^t,e_{k}^b;e_{k+1}^t,e_{k+1}^b)=
\langle e_{k+1}^b | \kappa_+(t_k-\tau_k/2;+1)| e^b_k\rangle 
\langle e_k^t| \kappa_-(t_k+\tau_k/2;-1)| e_{k+1}^t\rangle {\rm ~~ for ~~} b_k^+=+1; b_k^-=-1  
& & \nonumber\\
B(e_{k}^t,e_{k}^b;e_{k+1}^t,e_{k+1}^b)=
\langle e_{k+1}^b| \kappa_-(t_k+\tau_k/2;+1)| e_k^b\rangle 
\langle e_k^t|\kappa_+(t_k-\tau_k/2;-1) | e_{k+1}^t \rangle  {\rm ~~ for ~~} b_k^+=-1; b_k^-=+1
\end{eqnarray}

It is clear from Eqs. (\ref{s5e9}) that for equal $e_{k+1}^t=e_{k+1}^b=e_{k+1}$,
the expression $B(e_{k}^t,e_{k}^b;e_{k+1},e_{k+1})$ is nonzero only if $e_k^t=e_k^b$.
Hence, Eq. (\ref{s5e8}) can be represented as a matrix product

\begin{equation}
\label{s5e10}
\Pi_\kappa=\sum_{e_k}
\tilde I(e_1) \tilde B(e_1,e_2)\dots \tilde B(e_{N-2},e_{N-1})
\tilde B(e_{N-1},e_{N})\rho(e_N),
\end{equation}
where $\rho(e_N)=1/3$, $\tilde B(e_{k},e_{k+1})=B(e_{k},e_{k};e_{k+1},e_{k+1})$ and $\tilde I(e_1)=I(e_1,e_1)$. 
After the substitution of Eq. (\ref{s5e10}) in Eqs. (\ref{s5e0}) and (\ref{s5e6}),
each matrix element $\tilde B(e_{k},e_{k+1})$
multiplies by the correlation function 
$i^2\langle T_c b_k^-A_-(t_k+\tau_k/2;b_k^-) b_k^+A_+(t_k-\tau_k/2;b_k^+)\rangle$. The product
should be integrated over $d\tau_k$. One finally obtains

\begin{equation}
\label{s5e11}
I_{2N}=
\int_{-\infty}^0 dt_1 \int_{-\infty}^{t_1} dt_2 \dots 
\int_{-\infty}^{t_{N-2}} dt_{N-1}\langle I | \hat D^{N-1}|\rho\rangle,
\end{equation}
where 

\begin{equation}
\label{s5e12}
\langle\rho |=(1/3,~1/3,~1/3);
\end{equation}

\begin{eqnarray}
\label{s5e13}
\langle I| =-\frac{2e\nu}{\hbar^2}\sum_{k=1,2,3}
{\rm Re}\int_{-\infty}^0 dt 
\left[ 
\langle k|\kappa_+\kappa_-| k \rangle
{\frac{\exp(i\nu eVt/\hbar)\tau_c^{2\nu}}{(\delta+it)^{2\nu}}
} 
-
{\langle k| \kappa_-\kappa_+| k \rangle
\frac{\exp(i\nu eVt/\hbar)\tau_c^{2\nu}}{(\delta-it)^{2\nu}} }
\right]
\langle k| & & \nonumber\\
=-\frac{e\Gamma(1/3)\tau_c^{2/3}}{\sqrt{3}\hbar^2}
\left(\frac{eV}{3\hbar}\right)^{-1/3}\times(|C_3|^2,~|C_1|^2,~|C_2|^2);
\end{eqnarray}

\begin{equation}
\label{s5se14}
\hat D=-\frac{\sqrt{3}\Gamma(1/3)\tau_c^{2/3}}{\hbar^2}
\left(\frac{e V}{3\hbar}\right)^{-1/3}
\left (
\begin{array}{ccc}
|C_3|^2 & -|C_1|^2 & 0\\
0 &  |C_1|^2 & -|C_2|^2\\
-|C_3|^2 & 0 & |C_2|^2
\end{array}
\right ).
\end{equation}

The total current

\begin{equation}
\label{s5e15} 
I=\sum I_{2N}=\langle I |\exp(\bar t \hat D)|\rho\rangle,
\end{equation}
where $\bar t=+\infty$ is the length of the Keldysh contour.

All elements of the matrix $\hat D$ are real, all diagonal elements are negative, all nondiagonal elements are positive or zero and the sum of the elements in each column is zero. The Rohrbach theorem \cite{Rohrbach} applies to such matrices. Since $\langle \rho|\hat D=0$, $0$ is an eigenvalue of the matrix $\hat D$. According to the Rohrbach theorem, this eigenvalue is non-degenerate and the real parts of all other eigenvalues are negative. This allows for a
simple calculation of $\exp(\bar t\hat D)$. Let $\hat S$ be such a matrix that $\hat{\tilde D} =\hat S\hat D\hat S^{-1}$ 
assumes the Jordan normal form. Without
the loss of generality we can assume that the first column and the first string of the matrix $\hat{\tilde D}$ are zero. Then the first string of the matrix $\hat S$ can be chosen in the form $(1,~1,~1)$. Thus, the matrix exponent 

\begin{equation}
\label{s5e16}
\exp(\bar t\hat D)=\hat S^{-1}
\left (
\begin{array}{ccc}
1 & 0 & 0\\
0 & 0 & 0\\
0       & 0 & 0
\end{array}
\right ) \hat S=\hat S^{-1}
\left (
\begin{array}{ccc}
1 & 1 & 1 \\
0      & 0 & 0 \\
0       & 0 & 0
\end{array}
\right )
\end{equation}
Hence,

\begin{equation}
\label{s5e17}
I=\langle I | \hat S^{-1}| 1\rangle=\langle I|S^{-1}_1\rangle,
\end{equation}
where $|S^{-1}_1\rangle$ denotes the first column of the matrix $\hat S^{-1}$.

In order to complete our calculation we have to determine the components $f_1$, $f_2$, and $f_3$ of the vector $|S^{-1}_1\rangle$.
From the condition $\hat S\hat S^{-1}=\hat E$ one finds that 

\begin{equation}
\label{s5e18}
f_1+f_2+f_3=1.
\end{equation}
We also know that $|S^{-1}_1\rangle$ is the eigenvector of $\hat D$ with zero eigenvalue, 
i.e. $\hat D|S^{-1}_1\rangle=0$. Hence, for each $k$,

\begin{equation}
\label{s5e19}
|C_{k-1}|^2 f_{k}=|C_k|^2 f_{k+1},
\end{equation}
 where we use the convention $3+1=1$. The solution of the above equation
is $f_k=\alpha/|C_{k-1}|^2$, where $\alpha=\frac{1}{1/|C_1|^2+1/|C_2|^2+1/|C_3|^2}$ can be found from 
Eq. (\ref{s5e18}). Finally,

\begin{equation}
\label{s5e20}
I=\frac{-e\Gamma(1/3)\tau_c^{2/3}}{\sqrt{3}\hbar^2}\left(\frac{eV}{3\hbar}\right)^{-1/3}
\frac{3}{\frac{1}{|C_1|^2}+\frac{1}{|C_2|^2}+\frac{1}{|C_3|^2}}.
\end{equation}
This expression is equal to the harmonic average of the three tunneling currents
in three systems with a single quantum point contact with the tunneling amplitude $C_1$, $C_2$ and $C_3$ respectively. In other words, Eq. (\ref{s5e20}) is equivalent to (\ref{s2e4}). Using the result of Appendix C,
Eq. (\ref{s5e20}) can be represented as

\begin{equation}
\label{s5e21}
I=\frac{-e\Gamma(1/3)\tau_c^{2/3}}{\sqrt{3}\hbar^2}\left(\frac{eV}{3\hbar}\right)^{-1/3}
\left [
|\Gamma_1|^6+|\Gamma_2|^6+2|\Gamma_1\Gamma_2|^3\cos(3\alpha_0+2\pi\Phi/\Phi_0)
\right ]
\frac{|\Gamma_1|^2-|\Gamma_2|^2}{|\Gamma_1|^6-|\Gamma_2|^6},
\end{equation}
where $\alpha_0={\rm arg}[\Gamma_2/\Gamma_1]$. This allows one to easily verify Eq. 
(\ref{s2e7}) in the case $\nu=1/3$, $T=0$, $eVL,eVa\ll\hbar v$.

\subsection{Tunneling current for arbitrary filling factors, temperatures and voltages}

The calculations follow the same route as in the previous subsection.
The contribution to the current of order $2N$, $I_{2N}$, expresses via the trace $\Pi$ of the time-ordered product of $2N$ tunneling operators which form $N$ dipoles.
The trace must be integrated over the size of each dipole and the positions of $N-1$ dipoles, the remaining dipole being located at $t=0$. The trace 
$\Pi$ factorizes
as the product $\Pi=\Pi_\kappa\Pi_\phi$, where $\Pi_\kappa$ stays for the trace of the product of the Klein factors 
$\kappa_1$, $\kappa_2$, $\kappa_1^+$ and $\kappa_2^+$; 
$\Pi_\phi$ stays for the trace of the product of the operators $\exp(\pm i\sqrt{\nu}[\phi_1(x=0)-\phi_2(x=0)])$ and $\exp(\pm i\sqrt{\nu}[\phi_1(L)-\phi_2(L+a)])$. The latter trace factorizes  in the product of the two-point correlation functions \cite{bos} corresponding to each dipole: 

\begin{eqnarray}
\label{s5efg2}
F(b,c,t_1-t_2)={\rm Tr}[\hat\rho \exp(i\sqrt{\nu}\phi_1(x=b,t_1))\exp(-i\sqrt{\nu}\phi_1(x=0,t_2))]
{\rm Tr}[\hat\rho \exp(i\sqrt{\nu}\phi_2(x=c,t_1))\exp(-i\sqrt{\nu}\phi_2(x=0,t_2))]=
& & \nonumber \\
\left[\frac{\pi k_B T \tau_c/\hbar}{\sin(\pi k_B T[\delta+i(t_1-t_2-b/v)]/\hbar)}\right]^{\nu}
\left[\frac{\pi k_B T \tau_c/\hbar}{\sin(\pi k_B T[\delta+i(t_1-t_2-c/v)]/\hbar)}\right]^{\nu}
\end{eqnarray}
The former trace $\Pi_\kappa$ can be represented in the form similar to Eq. (\ref{s5e10}), where $\tilde B(e_k,e_{k+1})$ express via matrix elements of 
$\kappa_1$ and $\kappa_2$. Next, one can rewrite $I_{2N}$ in the form 
(\ref{s5e11}) with modified definitions of $|\rho\rangle$, $\langle I|$
and $\hat D$. $|\rho\rangle$ and $| I\rangle$ are now $1/\nu$-dimensional vectors;
$\hat D$ is a matrix of size $1/\nu\times 1/\nu$. 
Similar to the previous subsection, in order to obtain the matrix elements of $\hat D$, one has to multiply 
$\tilde B(e_k,e_{k+1})$ by the correlation function 
(\ref{s5efg2}) describing the dipole located at $t=t_k$ and then integrate the product over the dipole size.
One finds

\begin{equation}
\label{s5e23}
\langle \rho |=(\nu, \nu, \dots, \nu);
\end{equation}

\begin{equation}
\label{s5e24}
\langle I | e_k \rangle = e\nu (I_{k(V)}-I_{k(-V)});
\end{equation}

\begin{equation}
\label{s5e25}
\langle e_l|\hat D| e_k\rangle=\delta_{k,l} (I_{k(-V)}+I_{k(V)}) -
\delta_{k,l-1}I_{k(-V)} - \delta_{k,l+1}I_{k(V)},
\end{equation}
where

\begin{equation}
\label{s5e26}
I_{k(-V)}=\frac{1}{\hbar^2}\left[(|\Gamma_1|^2+|\Gamma_2|^2)j(-V;0)+
\Gamma_1\Gamma_2^*(\psi^*)^k j(-V;-a)+\Gamma_1^*\Gamma_2\psi^k j(-V;a)\right];
\end{equation}

\begin{equation}
\label{s5e27}
I_{k(V)}=\frac{1}{\hbar^2}\left[(|\Gamma_1|^2+|\Gamma_2|^2)j(V;0)+
\Gamma_1\Gamma_2^*(\psi^*)^{k-1} j(V;a)+\Gamma_1^*\Gamma_2\psi^{k-1} j(V;-a)\right],
\end{equation}
where $\psi=\exp(-2\pi i\nu)$, the phase factor $\exp(2\pi i \nu\Phi/\Phi_0+ie\nu VL/\hbar v)$ 
should be included in $\Gamma_2$ and

\begin{eqnarray}
\label{s5e28}
j(U,0)=-
\int_{-\infty}^{+\infty}dt\exp(i\nu eUt/\hbar)F(0,0,t); & & \nonumber\\
j(U,\pm a)=-\int_{-\infty}^{+\infty}dt\exp(i\nu eUt/\hbar) 
F(\pm L,\pm (L+a),t)
\end{eqnarray}

The total current is given by Eq. (\ref{s5e15}) of the previous section
with the above definitions of $\langle I|$, $\hat D$ and $|\rho\rangle$.
 
One can easily check that $j(U,x)=j^*(U,-x)$. Hence, $I_{k(\pm V)}$ are real.
Appendix D shows that

\begin{equation}
\label{s5e29}
I_{k(-V)}=\exp\left(-\frac{\nu eV}{k_B T}\right)I_{(k+1)(V)}=\gamma I_{(k+1)(V)}
\end{equation}
that is equivalent to the detailed balance condition (\ref{s2e8}).
A comparison with the geometry Fig. 3, Ref. \onlinecite{cfksw}, shows that the same integral which expresses the current $I_{\rm Fig. 3}$ in the geometry Fig. 3
also equals the expression 

\begin{equation}
\label{s5e30}
I_{k(V)}-I_{(k-1)(-V)}=(1-\gamma)I_{k(V)}=I^{(k)}_{\rm Fig. 3}/(\nu e)=-\frac{2^{2g}\tau_c}{\hbar^2}
|\Gamma_{\rm eff}|^2 \left(\frac{\pi k_B T \tau_c}{\hbar} \right)^{2\nu-1}
\frac{|\Gamma(\nu+\frac{i\nu eV}{2\pi k_B T})|^2}{\Gamma(2\nu)}\sinh\frac{\nu eV}{2k_B T},
\end{equation}
where

\begin{eqnarray}
\label{s5e31}
|\Gamma_{\rm eff}|^2=2\pi\frac{\Gamma(2\nu)}{\Gamma(\nu)}\sum_{m,l}\Gamma_m\Gamma_l^*
\exp({i[2\pi\nu\Phi/\Phi_0-e\nu V a/(2\hbar v)-2\pi (k-1)\nu][\delta_{m,2}-\delta_{l,2}]})\times & &
\nonumber\\ 
\frac{\exp(-\nu\pi a k_B T/[\hbar v])}{\sinh (eV\nu/2k_B T)}{\rm Im}
\left\{
\frac{\exp(ie\nu  V a/[2\hbar v])F(\nu,\nu-ie\nu V/2\pi k_B T; 1- ie\nu V/2\pi k_B T;e^{-2\pi k_B T a/[\hbar v]})}
{\Gamma(\nu+ie\nu V/2\pi k_B T)\Gamma(1-i e\nu V/2\pi k_B T)}
\right\},
\end{eqnarray}
$F$ being the hypergeometric function \cite{cfksw}. 

Since the particle current flows from the edge with the higher chemical potential to the edge with the lower potential, we know the sign of $I^{(k)}_{\rm Fig. 3}=dQ_1/dt>0$
(electron charge $e<0$). 
Hence, $I_{k(\pm V)}<0$. One can also see that the sum of the elements in any column of the matrix $\hat D$, Eq. (\ref{s5e25}),
is zero. Thus, the Rohrbach theorem \cite{Rohrbach} applies again and 
hence $\exp(\bar t \hat D)=\hat S^{-1}\hat M \hat S$, where the matrix $\hat S$ reduces $\hat D$
to the Jordan normal form, the matrix
$\hat M$ has only one non-zero element in its upper left corner and this element is equal to 1 just like in the preceding subsection. All elements of the first string of the matrix $\hat S$  
are equal to 1. This allows us to obtain the expression for the current in the form (\ref{s5e17}) following exactly the same steps as in section V.B.

The only thing left is the calculation of the components $f_r$ of the vector $|S^{-1}_1\rangle$, i.e. the first column of the matrix $\hat S^{-1}$.
From the condition $\hat S \hat S^{-1}=E$ one finds

\begin{equation}
\label{s5e32}
\sum_{r=1}^{1/\nu}f_r=1.
\end{equation}
From the condition $\hat D|S^{-1}_1\rangle=0$ one finds

\begin{equation}
\label{s5e33}
{I_{k(V)}f_k}-{I_{(k-1)(-V)}f_{k-1}}=I_{(k+1)(V)}f_{k+1}-I_{k(-V)}f_k.
\end{equation}
At the same time the current (\ref{s5e24},\ref{s5e17}) is

\begin{equation}
\label{s5e34}
I=\nu e \sum_k (I_{k(V)}-I_{k(-V)})f_k
\end{equation}

The system of equations (\ref{s5e32}-\ref{s5e34}) together with the detailed
balance condition (\ref{s5e29}) is equivalent to the system (\ref{s2e8}-\ref{s2e10}) of section II with $I_{k(\pm V)}$ playing the role of the transition probabilities $p$. The components $f_r$ of $|S^{-1}_1\rangle$
have the physical meaning of the distribution function. Thus, we can directly use the solution (\ref{s2e14}):

\begin{equation}
\label{s5e35}
I=\frac{1/\nu}{\sum_{k=1}^{1/\nu}I'_k[V,T,a]},
\end{equation}
where the current $I'_k=I^{(k)}_{\rm Fig. 3}$, Eq. (\ref{s5e30}), equals the tunneling current in the geometry Fig. 3 with the tunneling amplitudes
$\Gamma_1$ and $\Gamma_2\exp({i[2\pi\nu\Phi/\Phi_0-e\nu V a/(2\hbar v)-2\pi (k-1) \nu]})$.

In the limit $T=0; aeV\ll hv$ the above result reduces to a simple generalization of Eq. (\ref{s5e21}):

\begin{equation}
\label{B9}
I=\frac{-2e\nu\sin{(2\pi\nu)}\Gamma(1-2\nu)\tau_c^{2\nu}}{\hbar^2}\left(
\frac{eV\nu}{\hbar}
\right)^{2\nu-1}\frac{|\Gamma_1|^2-|\Gamma_2|^2}{|\Gamma_1|^{2/\nu}-|\Gamma_2|^{2/\nu}}
\left [
|\Gamma_1|^{2/\nu}+|\Gamma_2|^{2/\nu}
+2|\Gamma_1|^{1/\nu}|\Gamma_2|^{1/\nu}\cos (2\pi\Phi/\Phi_0+\alpha_0/\nu)
\right ].
\end{equation}
If $|\Gamma_1|\ne|\Gamma_2|$ the current never vanishes. At $|\Gamma_1|=|\Gamma_2|$ a ``resonance" is reached when 
$\Phi=[n+1/2-\alpha_0/(2\pi\nu)]\Phi_0$ and $I=0$.

\section{Conclusions}

We have found that the tunneling current through the electronic Mach-Zehnder interferometer is 
a periodic function of the magnetic flux with period $\Phi_0$. This result is valid both in the weak quasiparticle tunneling regime and in the weak electron tunneling regime. 
The relations between the flux-dependent and flux-independent components of the current, $I_\Phi$ 
and $I_0$, are different 
in these respective regimes. In the electron tunneling case, 
$I_\Phi(\Gamma_1,\Gamma_2)\sim[I_0(\Gamma_1,\Gamma_2)-I_0(\Gamma_1,0)]^{1/2}$ at low voltages and temperatures, 
$\Gamma_2\ll\Gamma_1$.
In the quasiparticle tunneling case the flux-dependent contribution scales as 
$I_\Phi(\Gamma_1,\Gamma_2)\sim[I_0(\Gamma_1,\Gamma_2)-I_0(\Gamma_1,0)]^{b}$, where $b>1$. 
The exponent in this power law
contains information about quasiparticle statistics since the exponent derives from the algebra of the Klein factors. 
Thus, the Mach-Zehnder interferometer
can be used to probe fractional statistics. Recently an interference pattern has been observed 
experimentally for the integer quantum Hall case in a Mach-Zehnder interferometer \cite{MZ}. 
Higher magnetic fields would allow an investigation of a 
fractional quantum Hall liquid.

\acknowledgments
We thank J. Chalker, M. P. A. Fisher, L. I. Glazman, M. Heiblum, I. Neder, X.-G. Wen and P. B. Wiegmann for useful discussions.
DEF thanks KITP where a part of this work was completed for hospitality. This research was supported 
in part by the National Science Foundation under Grant No. PHY99-07949, by the U.S.-Israel Binational Science Foundation, and the ISF of the Israel Academy of Sciences.

\appendix

\section{Electron tunneling at low voltages and zero temperature}

In this appendix we verify that the current, Eqs. (\ref{15}-\ref{15b}), satisfies the asymptotics (\ref{s4e1}) at 
$V\rightarrow 0$. The flux-independent contribution to the current (\ref{15a}) scales as $V^{g_1+g_2-1}$,
where $g_1=1/\nu_1$ and $g_2=1/\nu_2$. Hence, we need to check
that the flux-dependent contribution $I_\Phi$ does not exceed ${\rm const}V^{g_1+g_2-1}$ at low voltages.
Expanding $\exp(-\frac{ieVz}{\hbar})$ in (\ref{15b}) in powers of $ieVz/\hbar$ one finds that the coefficient before 
$V^k$ in the Taylor expansion of $I_{\Phi}$ equals

\begin{equation}
\label{A1}
s_k={\rm const}\left [ 
\frac{1}{(g_1-1)!}\frac{d^{g_1-1}}{dz^{g_1-1}}|_{z=0}\frac{z^k}{(z+a/v)^{g_2}}
+
\frac{1}{(g_2-1)!}\frac{d^{g_2-1}}{dz^{g_2-1}}|_{z=-\frac{a}{v}} z^{k-g_1}
\right ].
\end{equation}
We want to show that $s_k=0$ for all $k<g_1+g_2-1$.

Let us consider separately $k>g_1-1$ and $k\le g_1-1$.

1) $k>g_1-1$. This inequality can be satisfied for $g_2>1$ only since $k<g_1+g_2-1$.
In this case $\frac{d^{g_2-1}}{dz^{g_2-1}}|_{z=-\frac{a}{v}} z^{k-g_1}=0$ since $g_2-1>k-g_1\ge 0$.
Since
$\frac{d^p}{dz^p}|_{z=0}z^k=0$ at
$p<k$, one finds $\frac{d^{g_1-1}}{dz^{g_1-1}}|_{z=0}\frac{z^k}{(z+a/v)^{g_2}}=\sum_{p=0}^{g_1-1}C_{g_1-1}^p 
\frac{d^p}{dz^p}z^k\frac{d^{g_1-1-p}}{dz^{g_1-1-p}}\frac{1}{(z+a/v)^{g_2}}=0$, where $C_{a}^b$ 
denote binomial coefficients. Thus, $s_k=0$ for $g_1+g_2-1>k>g_1-1$.

2) At $k\le g_1-1$ one finds

\begin{equation}
\label{A2}
s_k={\rm const}\left[
\frac{1}{(g_1-1)!}C_{g_1-1}^k\frac{k! (-1)^{g_1-1-k}}{(a/v)^{g_1+g_2-k-1}}\frac{(g_1+g_2-k-2)!}{(g_2-1)!}+
\frac{1}{(g_2-1)!}(-a/v)^{k-g_1-g_2+1}\frac{(g_1+g_2-k-2)!}{(g_1-k-1)!}
\right]=0.
\end{equation}

Thus, $s_k=0$ for all $k<g_1+g_2-1$.

\section{Perturbation theory at finite temperature}

We want to calculate the current, Eq. (\ref{12}), at a finite temperature, $k_B T>0$. The correlation function,
Eq. (\ref{13}),
is \cite{bos}

\begin{equation}
\label{C1}
F(b,c,t)=\left[\frac{\pi k_B T \tau_c/\hbar}{\sin(\pi k_B T[\delta+i(t-b/v)]/\hbar)}\right]^{1/\nu_1}
\left[\frac{\pi k_B T \tau_c/\hbar}{\sin(\pi T[\delta+i(t-c/v)]/\hbar)}\right]^{1/\nu_2}.
\end{equation}

Since $1/\nu_1=g_1$ and $1/\nu_2=g_2$ are odd integers, 
 $F(-L,-(L+a),t)=F(L,L+a,-t)$ on the real axis except in the vicinity of the two real poles
$t=L/v$ and $t=(L+a)/v$. $F(0,0,t)=F(0,0,-t)$ except near $t=0$. Hence, the integral, Eq. (\ref{12}), reduces to the sum of the contour integrals along small circles around the real poles. A straightforward calculation yields

\begin{equation}
\label{C2}
I=\frac{e\tau_c}{\hbar^2}\left(\frac{\pi k_B T \tau_c}{\hbar}\right)^{g-1}(|\Gamma_1|^2+|\Gamma_2|^2)J_0 + 
\frac{e\tau_c}{\hbar^2}\left(\frac{\pi k_B T \tau_c}{\hbar}\right)^{g-1}[\Gamma_1\Gamma_2^*
\exp(-2\pi i\Phi/\Phi_0) J_\Phi+c.c.],
\end{equation}

where $g=g_1+g_2=1/\nu_1+1/\nu_2$, and

\begin{equation}
\label{C3}
J_0=\frac{2\pi}{i^{g+1}(g-1)!}\frac{d^{g-1}}{dz^{g-1}}|_{z=0}\frac{\exp(ieVz/\pi k_B T)z^g}{\sinh^g z},
\end{equation}

\begin{equation}
\label{C4}
J_\Phi=\frac{2\pi}{i^{g+1}(g_1-1)!}\frac{d^{g_1-1}}{dz^{g_1-1}}|_{z=0}
\frac{\exp(\frac{ieVz}{\pi k_B T})z^{g_1}}{(\sinh z)^{g_1}(\sinh[z-\frac{\pi a k_B T}{\hbar v}])^{g_2}}
+\frac{2\pi}{i^{g+1}(g_2-1)!}\frac{d^{g_2-1}}{dz^{g_2-1}}|_{z=\frac{\pi a k_B T}{\hbar v}}
\frac{\exp(\frac{ieVz}{\pi k_B T})(z-\frac{\pi a k_B T}{\hbar v})^{g_2}}{(\sinh z)^{g_1}(\sinh[z-\frac{\pi a k_B T}{\hbar v}])^{g_2}}
\end{equation}

Note that the temperature enters the above expression in the combination $ak_B T$ but not in the combination $Lk_B T$.
This can be understood from the picture of non-interacting electrons. For non-interacting particles the current 
is the sum of independent contributions from different electron energies. Each contribution depends on the phase difference between the two paths connecting the point contacts but not the total phase accumulated on any of those paths. The former is proportional to $a$, the latter to $L$.

One can easily extract the asymptotical behavior of the linear conductance $G=\frac{dI}{dV}$ at low temperatures (\ref{16})
from equations (\ref{C3},\ref{C4}). In the Loran expansion of $G$ in powers of $a$, each term of order $a^n$ 
is proportional to $T^{1/\nu_1+1/\nu_2-2+n}$ since $a$ enters the expression for the current in the combination 
$(ak_B T/\hbar v)$ only. In the limit $a\rightarrow 0$ the conductance must remain finite. Indeed, 
at $L=0$ this limit corresponds to a problem with a single tunneling contact. Hence, only positive and zero powers of $a$ are present in the Loran expansion and the leading contribution to the temperature dependence of the conductance scales as 
$T^{1/\nu_1+1/\nu_2-2}$.

\section{Effective single impurity model}

In the effective single impurity model, section V.B, the current is given by Eq. (\ref{s5e20}). The purpose of this appendix consists in the calculation of the
coefficient in Eq. (\ref{s5e20}),

\begin{equation}
\label{B1}
U=\frac{N}{\sum_{k=1}^{N}\frac{1}{|C_k|^2}},
\end{equation}
where $C_k=\Gamma_1+\Gamma_2\exp(2\pi i\Phi/[N\Phi_0]- 2\pi i k/N)$, Eq. (\ref{s5e5}), $N=1/\nu$. 
Without the loss of generality we can assume that $\Gamma_1=|\Gamma_1|$ is
real. Let $\Gamma_2=|\Gamma_2|\exp(i\alpha_0)$. In the rest of the appendix we will use the notation $\gamma_1=|\Gamma_2|$, $\gamma_2=|\Gamma_2|$ and
$\phi=2\pi\Phi/[N\Phi_0]+\alpha_0$. 
$U$ can be represented as the ratio of two polynomials of $\gamma_1$ and $\gamma_2$:

\begin{equation}
\label{B2}
U=N\frac{\Pi_k|\gamma_1+\gamma_2\exp(i\phi+2\pi i k/N)|^2}
{\sum_k\Pi'_{l\ne k}|\gamma_1+\gamma_2\exp(i\phi+2\pi i l/N )|^2},
\end{equation}
where the prime after the product sign means that the term with $l=k$ is
not included in the product. The nominator in Eq. (\ref{B2}) equals $|P(\gamma_2\exp(i\phi))|^2$, where the polynomial

\begin{equation}
\label{B3}
P(z)=\Pi_k[\gamma_1+z\exp(2\pi i k/N)]
\end{equation}
All roots of the polynomial (\ref{B3}) coincide with the roots of the polynomial
$\gamma_1^N+z^N$. Hence, from the basic theorem of algebra 

\begin{equation}
\label{B4}
P(z)=\gamma_1^N+z^N
\end{equation}
and the nominator is

\begin{equation}
\label{B5}
|P(\gamma_2\exp(i\phi))|^2=(\gamma_1^N+\gamma_2^N\exp(iN\phi))
(\gamma_1^N+\gamma_2^N\exp(-iN\phi))=\gamma_1^{2N}+\gamma_2^{2N}
+2\gamma_1^N\gamma_2^N\cos N\phi.
\end{equation}

The denominator in Eq. (\ref{B2}) can be represented as

\begin{eqnarray}
\label{B6}
d=\sum_k \left |
\frac{P(\gamma_2\exp(i\phi))}{\gamma_1+\gamma_2\exp(i\phi+2\pi i k/N)}
\right | ^2 = 
\sum_k |\sum_{p=0}^{N-1} (-1)^p\gamma_1^{N-1-p}\gamma_2^{p}\exp(ip\phi+2\pi pki/N) |^2 =
& & \nonumber\\
\sum_{k=0}^{N-1}\sum_{p=0}^{N-1}\sum_{r=0}^{N-1}
\gamma_1^{N-1-p}\gamma_2^p(-1)^p\exp(ip\phi+2\pi kpi/N)\times
\gamma_1^{N-1-r}\gamma_2^r(-1)^r\exp(-ir\phi-2\pi kri/N).
\end{eqnarray}
The sum $\sum_{k=0}^{N-1}\exp(2\pi k(p-r)i/N)=0$, if $p-r\ne nN$. With the help of this property, Eq. (\ref{B6}) reduces to

\begin{equation}
\label{B7}
d=N\sum_{p=0}^{N-1}\gamma_1^{2(N-1)-2p}\gamma_2^{2p}=
N\frac{\gamma_1^{2N}-\gamma_2^{2N}}{\gamma_1^2-\gamma_2^2}.
\end{equation}

Finally, the combination of Eqs. (\ref{B5}) and (\ref{B7}) yields

\begin{equation}
\label{B8}
U=\frac{\gamma_1^2-\gamma_2^2}{\gamma_1^{2N}-\gamma_2^{2N}}
\left [
\gamma_1^{2N}+\gamma_2^{2N}
+2\gamma_1^N\gamma_2^N\cos N\phi
\right ]
\end{equation}

\section{Detailed balance}

Here we derive Eq. (\ref{s5e29}). We need to show that 

\begin{equation}
\label{D1}
j(-V;0)=\gamma j(V;0); j(-V;\pm a)=\gamma j(V;\mp a)
\end{equation}
This is equivalent to the equation

\begin{equation}
\label{D2}
\int_{-\infty}^{+\infty}dt F(-b,-c,t)\exp(-\frac{ie\nu Vt}{\hbar})=
\gamma\int_{-\infty}^{+\infty}dt F(b,c,t)\exp(\frac{ie\nu Vt}{\hbar}),
\end{equation}
where $F$ is given by Eq. (\ref{s5efg2}).
Both integrals are taken over contour a) which
goes below the real axis in Fig. 8. We can change the sign of $t$
in the first integral in Eq. (\ref{D2}). This also changes the integration contour into
contour b), Fig. 8 (infinitesimal $\delta$
is positive). The integral over contour b) is equal to the integral over contour c). The latter integral
equals $\gamma\int_{-\infty}^{+\infty}dt F(b,c,t)\exp(\frac{ie\nu Vt}{\hbar})$ indeed.

\newpage

\begin{figure}
\epsfig{file= 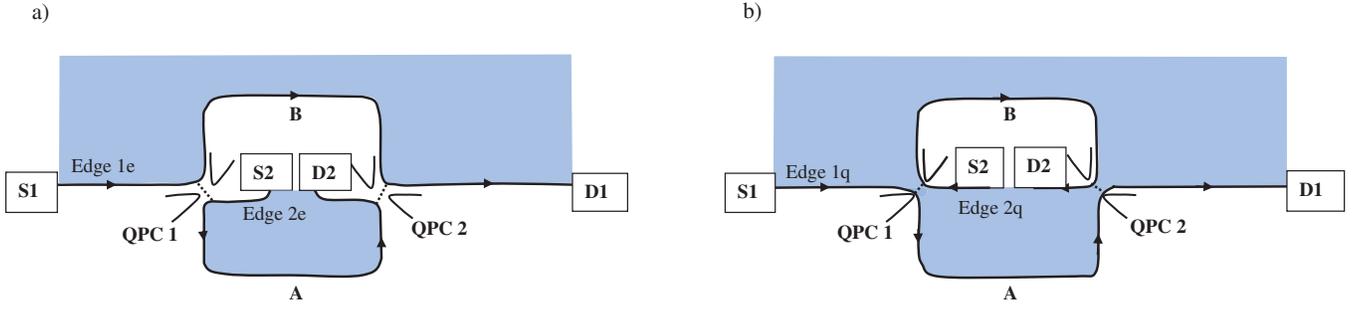 
, width=\linewidth} 
  \caption{Schematic picture of the Mach-Zehnder interferometer.
S and D denote sources and drains. Arrows indicate the propagation direction of the chiral edge modes. 
The source and drain voltages are $V_{S1}=V$, $V_{D1}=V_{D2}=V_{S2}=0$.
a)  Electrons tunnel between the puddle defined by edge 2e and the QHE strip (edge 1e)
b) Fractionally charged quasiparticles tunnel between edges 1q and 2q. In both cases the difference of the chemical potentials equals the voltage drop between S1 and S2.
    }
\label{fig1}
\end{figure} 

\begin{figure}
   \epsfig{file=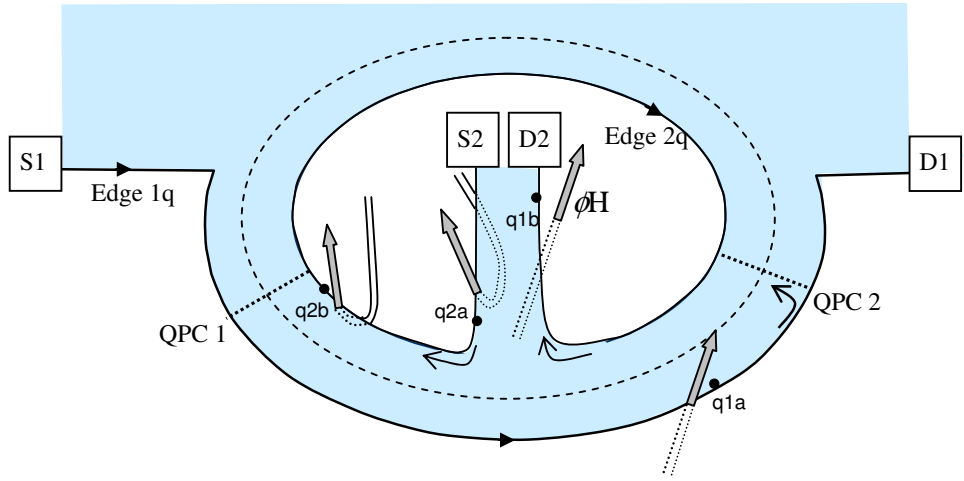   
, width=5in} 
  \caption{ Quasiparticle propagation and tunneling in the Mach-Zehnder interferometer.
Once quasiparticle $q_{1a}$ with its attached flux tube tunnels through QPC2 it arrives in D2 ($q_{1b}$) and the flux tube is released within the MZ hole. The flux tube $\delta \Phi$ affects the statistical phase accumulated by the next tunneling quasiparticle. A quasiparticle emitted from S2 will have its flux tube folded ($q_{2b}$) and will not affect the statistical phase accumulated by other quasiparticles.
    }
\label{fig3}
\end{figure}

\begin{figure}
\epsfig{file=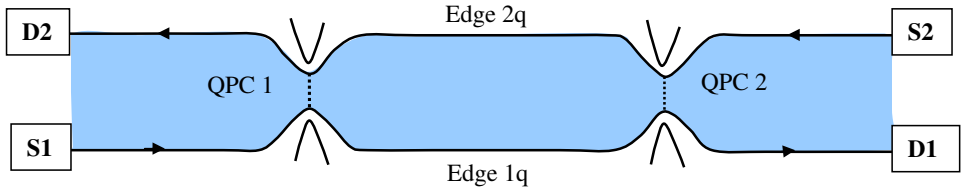  
, width=5in} 
  \caption{The Aharonov-Bohm interferometer from Ref. \onlinecite{cfksw}.
    }
\label{figX}
\end{figure}

\begin{figure}
\epsfig{file=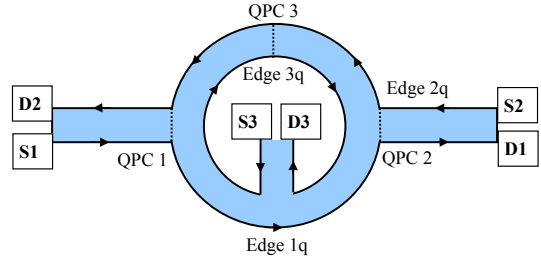 
, width=5in} 
  \caption{The Aharonov-Bohm interferometer considered in Ref. \onlinecite{kane}
    }
\label{figXX}
\end{figure} 

\begin{figure}
   \epsfig{file= 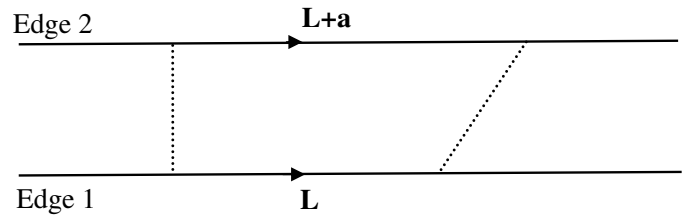
, width=3.5in} 
  \caption{ Mach-Zehnder interferometer can be modeled as two chiral Luttinger liquids with the same propagation direction. Dashed lines represent quantum point contacts. In the case corresponding to Fig. 1b) quasiparticle tunneling is 
allowed while in the case that corresponds to Fig. 1a) only electron tunneling is allowed. 
    }
\label{fig2}
\end{figure}

\begin{figure}
   \epsfig{file= 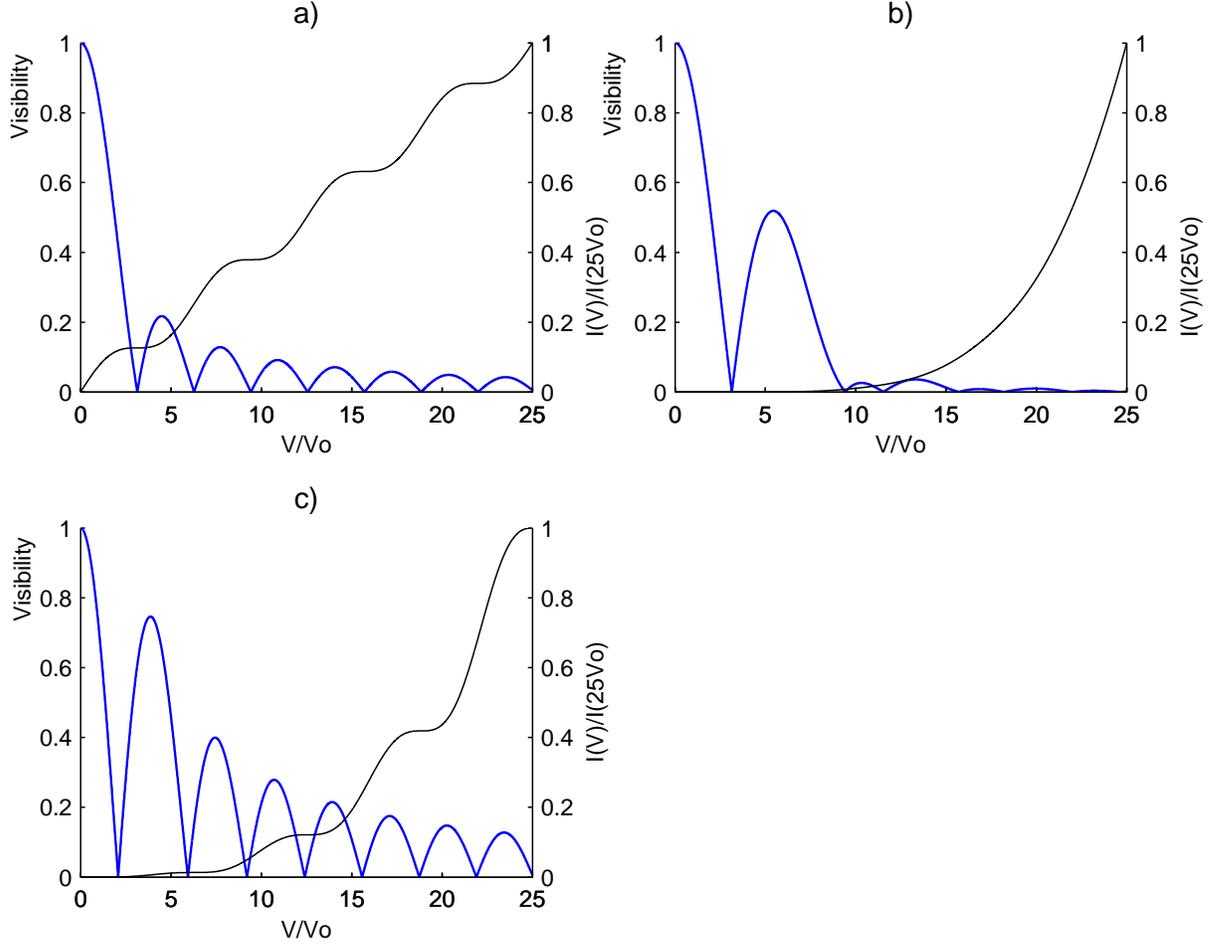
, width=\linewidth} 
  \caption{ Current $I(V,\Phi=0)$ (thin line) and visibility 
$R=[{\rm max}_\Phi I(V,\Phi)-{\rm min}_\Phi I(V,\Phi)]/\
[{\rm max}_\Phi I(V,\Phi)+{\rm min}_\Phi I(V,\Phi)]={\rm max} I_\Phi/I_0$ (thick line)
as functions of $V/V_0\equiv eVa/(\hbar v)$. The current is normalized to its value at $V=25 V_0$; $\Gamma_1=\Gamma_2$;
a) $\nu_1=\nu_2=1$; b) $\nu_1=1,\nu_2=1/3$; c) $\nu_1=\nu_2=1/3$. [cf. Refs. 
\onlinecite{cfksw,geller}.]} 
\label{fig4}
\end{figure} 

\begin{figure}
\epsfig{file=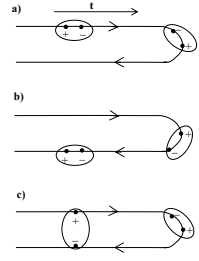 
, width=3in} 
  \caption{The 4th order contributions in the Keldysh technique can be represented in terms of two dipoles one of which is located at $t=0$. The second dipole can include a) two charges from the top branch of the contour; b) two charges from the bottom branch; or c) one charge from the top branch and one from the bottom branch.
    }
\label{figY}
\end{figure} 

\begin{figure}
\epsfig{file=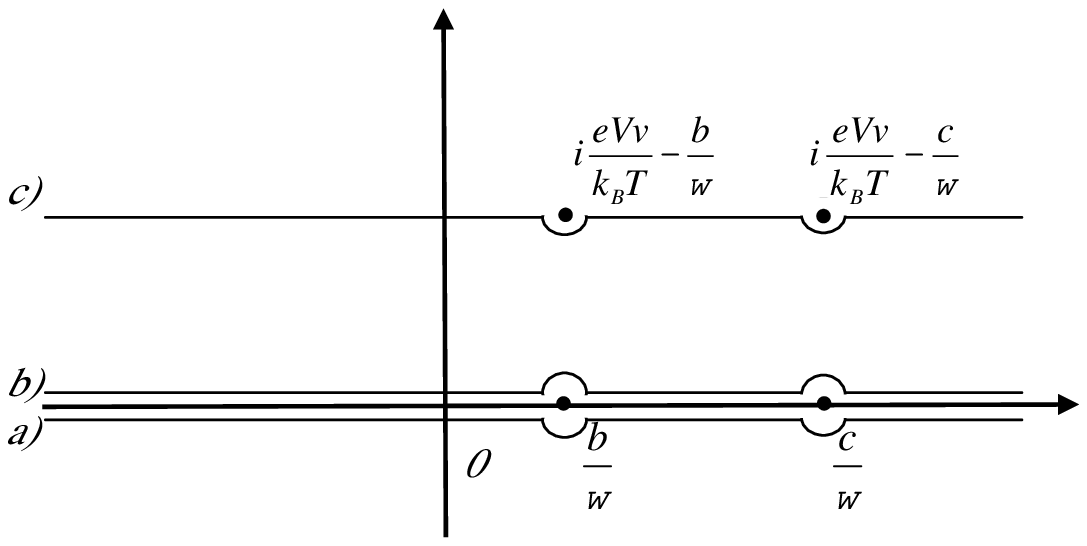 
, width=6in} 
  \caption{Three integration contours in the complex plane  which we use in the derivation of the detailed balance condition.
    }
\label{figA}
\end{figure} 


\begin{thebibliography} {150}
\bibitem{footnote}
The phase factor can be derived from the chiral Luttinger liquid model Eq. (\ref{1}) with the following argument. 
A small change of the vector potential $\delta \vec A$
leads to the additional contribution $\delta\vec A\vec j/c$ 
in the action, where $\vec j$ is the current. 
The current operator is proportional to the excitation velocity  $v$ (\ref{1}) and the charge density (\ref{3}). 
The density $\rho_l\sim\partial_x\phi_l=\partial_x[{\rm const}+N_lx+\sum_{k\ne 0} \exp(ikx)\phi_l(k)]$.
Thus, the additional contribution to the action is linear in the Bose fields. 
Hence, one can shift the Bose fields $\phi_l$ 
by functions of the coordinate in such a way that the vector potential term $\delta\vec A\vec j/c$ in the action gauges out. After this transformation a phase factor emerges in the tunneling operator.
\end{thebibliography}
\end{document}